\documentclass[
  journal=largetwo,
  manuscript=article-type,
  year=2020,
  volume=37,
]{cup-journal}

\usepackage{amsmath}
\usepackage[nopatch]{microtype}
\usepackage{booktabs}

\title{Evidence for the spin-kick alignment of pulsars from the statistics of their magnetic inclinations}

\author{Anton Biryukov}
\affiliation{The Raymond and Beverly Sackler School of Physics and Astronomy, Tel Aviv University, Tel Aviv, 6997801, Israel}
\alsoaffiliation{Sternberg Astronomical Institute, Lomonosov Moscow State University, 13 Universitetsky pr., Moscow, 119234, Russia}
\email[A. Biryukov]{ant.biryukov@gmail.com}

\author{Gregory Beskin}
\affiliation{Special Astrophysical Observatory, Nijniy Arkhyz, Karachaevo-Cherkessia, 369167, Russia}
 %% First letter not capped

\begin{document}

\begin{abstract}
It is thought that isolated neutron stars receive a natal kick velocity at birth nearly aligned with their spin axis. Direct observational confirmation of this alignment is currently limited to a single source in a supernova remnant (PSR J0538+2817), for which the three-dimensional velocity has been well constrained. Meanwhile, pulsar polarisation statistics suggest the existence of a spin-kick correlation, though both aligned and orthogonal cases remain possible. However, if the velocities of radiopulsars are predominantly aligned with their spin axes, a systematic difference in the observed transverse velocities of pulsars with small and large magnetic obliquities would be expected. In particular, due to projection effects, weakly oblique rotators should exhibit smaller, less scattered transverse velocities. Conversely, the transverse velocities of pulsars with large magnetic inclination should reflect their actual three-dimensional velocities. This study uses this idea to analyse samples of 13 weakly and 25 strongly oblique pulsars with known distances and proper motions. We find that their peculiar velocities are distributed differently, with statistical confidence levels of 0.007 and 0.016 according to the Anderson–Darling and Kolmogorov–Smirnov tests, respectively. We performed a detailed population synthesis of isolated pulsars, considering the evolution of their viewing geometry in isotropic and spin-aligned kick scenarios. The observed split in the transverse velocity distributions and its amplitude are consistent with the spin-aligned kick model, but not with the isotropic case. At the same time, an orthogonal kick would predict a similar effect, but with the opposite sign. This provides robust support for pulsar spin kick alignment based on statistics, independently of polarisation.
\end{abstract}

%\noindent

\section{Introduction}
The observed peculiar velocities of isolated radio pulsars with respect to the Galactic environment are $\sim 200-500$ km s$^{-1}$ \citep[e.g.][]{Igoshev20}, which is much faster than $\sim 15-40$ km s$^{-1}$ for their high-mass progenitors \citep[e.g.][]{Tetzlaff_HyperVel11,Carre_HyperVel23}. This suggests that neutron stars experience a significant natal kick during their formation in supernova explosions. Unfortunately, the absence of observational techniques to determine the full three-dimensional velocities of neutron stars has impeded the firm establishment of the statistical properties of their Galactic motion\footnote{See, however, the discussion in Sect. 2.5.3 by \cite{TEMPO2}}. Therefore, the details of the distribution of pulsar kicks remain somewhat uncertain. Thus, some researchers advocate for a broad unimodal velocity distribution \citep{fgk06, 2005MNRAS.360..974H, 2009ApJ...698..250C}, while others identify two components with typical dispersions of $\sim 100$ and $\sim 500$ km s$^{-1}$, respectively \citep{1998ApJ...505..315C, 2002ApJ...568..289A, Verbunt17}. The physical origin of such bimodality remains unclear. One widely discussed suggestion is that a significant fraction of isolated neutron stars could originate from disrupted binary systems \citep{kuranov09, chm10, Verbunt17}. Indeed, a supernova explosion could affect the orbital velocities of a binary system that survives the event \citep{2024ApJ...966...17B}. However, if the explosion is powerful enough to disrupt the system, the natal kick will overlap the orbital velocity. It has been shown that neutron stars born in disrupted systems likely have identical velocity distributions to those formed from isolated stars \citep{kuranov09}. Therefore, this work does not distinguish between the two origins of pulsars.

\begin{figure*}
\centering
\includegraphics[width=1.1\linewidth]{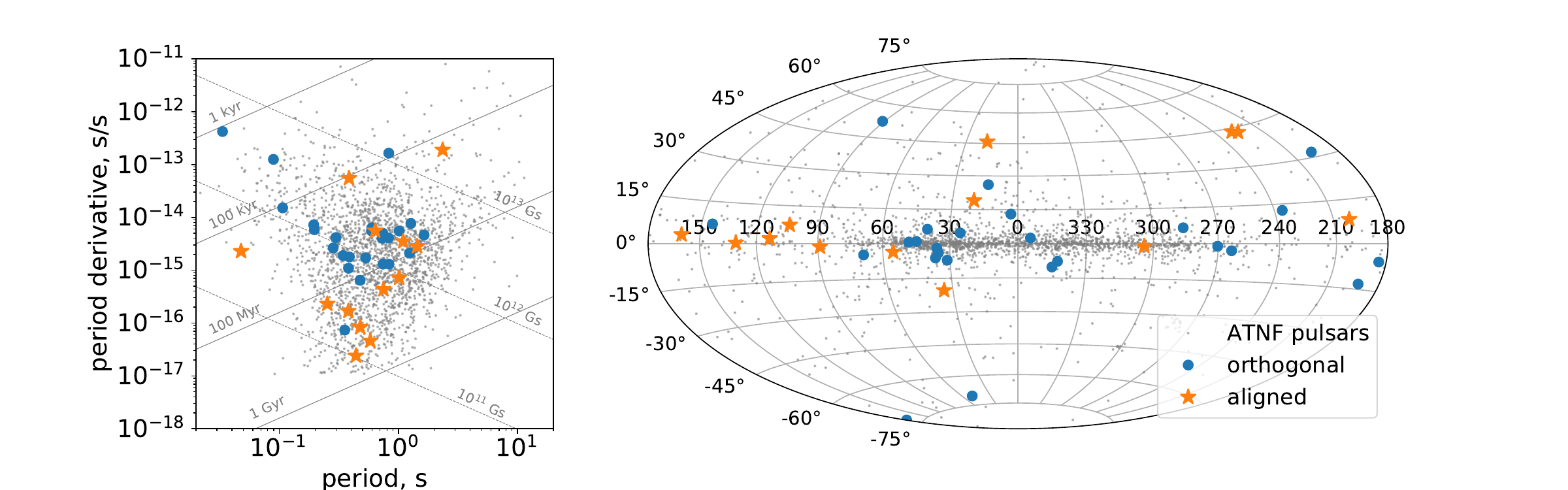}
\caption{Pulsars under consideration. {\bf Left plot}: the $P-\dot P$ pulsar diagram. Orange stars show weakly aligned rotators (with small magnetic angles). Blue circles show nearly orthogonal ones (with large magnetic angles). Grey dots represent classical (non-recycled) isolated pulsars listed in the ATNF catalogue. {\bf Right plot}: The same pulsars are plotted in the Galactic sky coordinates.}
\label{fig:pulsars_map}
\end{figure*}

At the same time, the kick velocities of neutron stars are thought to be related to their rapid spin-up at birth, as both processes could result from the interaction between the newborn star and the supernova ejecta \citep[e.g.][]{Wang07_SpinVel, Janka17, Janka22, 2022MNRAS.517.3938C}. Therefore, it makes sense to expect a correlation between the pulsar spin axis and velocity directions. This correlation should exist for the peculiar velocity of objects younger than $\sim 10$ Myr. However, it breaks down at Galactic dynamical timescales, as discussed by \cite{Mandel_Igoshev23}.

Studies of pulsar radio emission have shown that their apparent transverse velocities are distributed anisotropically with respect to their spin axes. Evidence of a strong correlation between spin and velocity was found \citep{Johnston05, Wang06_SpinVel, nout12, nout13}. This analysis was based on applying the Rotating Vector Model (RVM) to the pulsar's linear polarisation signal. This model assumes that the polarisation position angle follows the projection of a rotating dipole field onto the viewing plane \citep{RVM, RVM_Lyutikov}. However, this approach has two limitations. Firstly, two orthogonal polarisation modes are possible for electromagnetic waves propagating through a magnetic field. These X- and O- (extraordinary and ordinary, respectively) modes may be present in observations of a single pulsar and cannot be distinguished \citep{1984ApJS...55..247S, 1997A&A...327..155G, 2014MNRAS.441.1943W}. Secondly, the RVM is insensitive to the spin direction of a radio pulsar; a clockwise rotator exhibits the same polarisation behaviour as a counterclockwise rotator. Together, these degeneracies prevent us from determining whether pulsar velocities are aligned with or perpendicular to their spins. Both scenarios remain possible.

Nevertheless, \cite{Yao21} have recently presented the first clear detection of three-dimensional spin-velocity alignment in PSR J0538+2817. This is the only such evidence obtained so far. Therefore, alternative observational verification involving the statistics of a subset of objects, independent of polarisation, is necessary.

This work presents a method of testing spin-velocity alignment in isolated radio pulsars based on geometric considerations. If the spins of pulsars align with their velocities, the transverse components of the latter for objects with different magnetic angles must differ. Magnetically aligned pulsars (with small spin-magnetic axis angles) move almost along the line of sight, resulting in small, weakly scattered transverse velocities. In contrast, magnetically orthogonal pulsars move perpendicular to the line of sight, meaning their transverse velocities are close to their full 3D velocities. Consequently, one would expect different transverse velocity distributions for nearly aligned and orthogonal pulsars.

We tested this idea using existing data on the magnetic inclinations, proper motions and distances of isolated radio pulsars. Additionally, we performed a detailed population synthesis to verify that the observed difference in the velocity distribution was consistent with the properties of a realistic population, and could be detected in observations.

The paper is organised as follows. In Section~\ref{sec:trans_vels} we compare the transverse velocities of radio pulsars with different magnetic obliquities. Section~\ref{sect:popsynth} describes the population synthesis of these objects, with the results presented in Section~\ref{sect:results_main}. Section~\ref{sect:discuss} provides a brief discussion, while conclusions are collected in Section~\ref{sect:conclude}. \ref{app:pop_synth_results} contains the table with precise details of the pulsar subset under investigation and figures showing the expanded results of the population synthesis undertaken.

\section{Pulsars' velocity distributions}
\label{sec:trans_vels}

\subsection{Magnetically orthogonal and aligned pulsars}
In order to compare the transverse velocities of magnetically aligned and orthogonal pulsars, it is necessary to create representative subsets of observed sources for each alignment type. In our previous work \citep{bb23}, we compiled a catalogue of 77 isolated radio pulsars identified in the literature as having extreme magnetic inclination angles, either small ($\alpha$ near 0) or large ($\alpha$ near 90 degrees). The most reliable method of estimating $\alpha$ involves fitting the rotating vector model (RVM) to the pulsar's linear polarisation signal.  However, achieving high precision in this estimation often proves challenging, primarily due to the low signal-to-noise ratio in the outer ``wings'' of the average pulse profile. Indeed, polarisation data carry the most information about the magnetic inclination angle in this part of the pulse \citep{FAST_polarization23, Meerkat_RVM}. Therefore, the magnetic inclination angle of a particular pulsar is usually unknown.

However, the pulsar emission properties could provide valuable information for distinguishing between weakly and strongly oblique rotators without the need for precise angle estimation. These properties include average pulse widths, distinguishable core and cone emission components, the presence of an interpulse and qualitatively different polarisation behaviour in both the main pulse and interpulse. Many authors have used these properties to classify pulsars into two categories \citep{lyne88, rankin90, rankin93a, maciesiak11a, keith10, malov13, jk19}. The list of 77 pulsars mentioned above was compiled using these classification criteria. For the present study, we extracted a subset of this list, retaining only objects with determined full proper motions and distances. We found 13 nearly aligned and 25 almost orthogonal rotators that meet these conditions. Figure~\ref{fig:pulsars_map} plots these pulsars on the classical $P-\dot{P}$ diagram and within galactic coordinates, alongside other single, non-recycled pulsars from the ATNF database \citep{atnf}\footnote{v2.0.0}\footnote{ \texttt{https://www.atnf.csiro.au/research/pulsar/psrcat/}}. Despite the limited sample size, the selected pulsars provide a representative cross-section of the overall pulsar population. Details of the selected pulsars are presented in Table~\ref{tab:pulsars} in \ref{app:pop_synth_results}. The parameters of these objects were also taken from the ATNF. Half the distances were estimated from dispersion measures, while parallaxes were determined independently for 19 objects.

\subsection{Pulsars' peculiar velocities}
Consider the motion of pulsars relative to the galactic inertial reference frame $(x, y, z)$, with its origin at the Solar System Barycentre. The $x$-axis is directed towards the galactic centre, the $y$-axis is aligned with the galactic rotation and the $z$-axis points towards the north galactic pole. If $l$ and $b$ are the pulsar's galactic coordinates and $\mu_l$ and $\mu_b$ are the corresponding proper motions, with $d$ being the distance to the pulsar, then the apparent transverse velocity vector of the pulsar is
\begin{equation}
    \mathbf{v}_{\mathrm t} = kd\mu_l\left ( 
    \begin{array}{c}
    -\sin l \cos b \\
    \cos l \cos b \\
    0
    \end{array} 
    \right) + kd \mu_b\left ( 
    \begin{array}{c}
    -\cos l \sin b \\
    -\sin l \sin b \\
    \cos b
    \end{array} 
    \right),
\end{equation}
where $k \approx 4.74$~km~s$^{-1}$/(pc~arcsec~yr$^{-1}$). Physically, this velocity comprises the Sun's Galactic motion, the circular velocity of a pulsar's local standard of rest (LSR) within the Galaxy and the peculiar velocity ${\mathbf v}_{pec,t}$ relative to the LSR. We assume that only the latter component generally contains information about the kick that a pulsar received after a supernova explosion. Therefore, it relates to the kick-velocity correlation. Although this correlation spreads out over a pulsar's lifetime \citep{Mandel_Igoshev23}, it still applies to relatively young objects that have not yet crossed their death line (see also Section~\ref{sect:popsynth_results} below).

\begin{figure}[hbt!]
\centering
\includegraphics[width=1\linewidth]{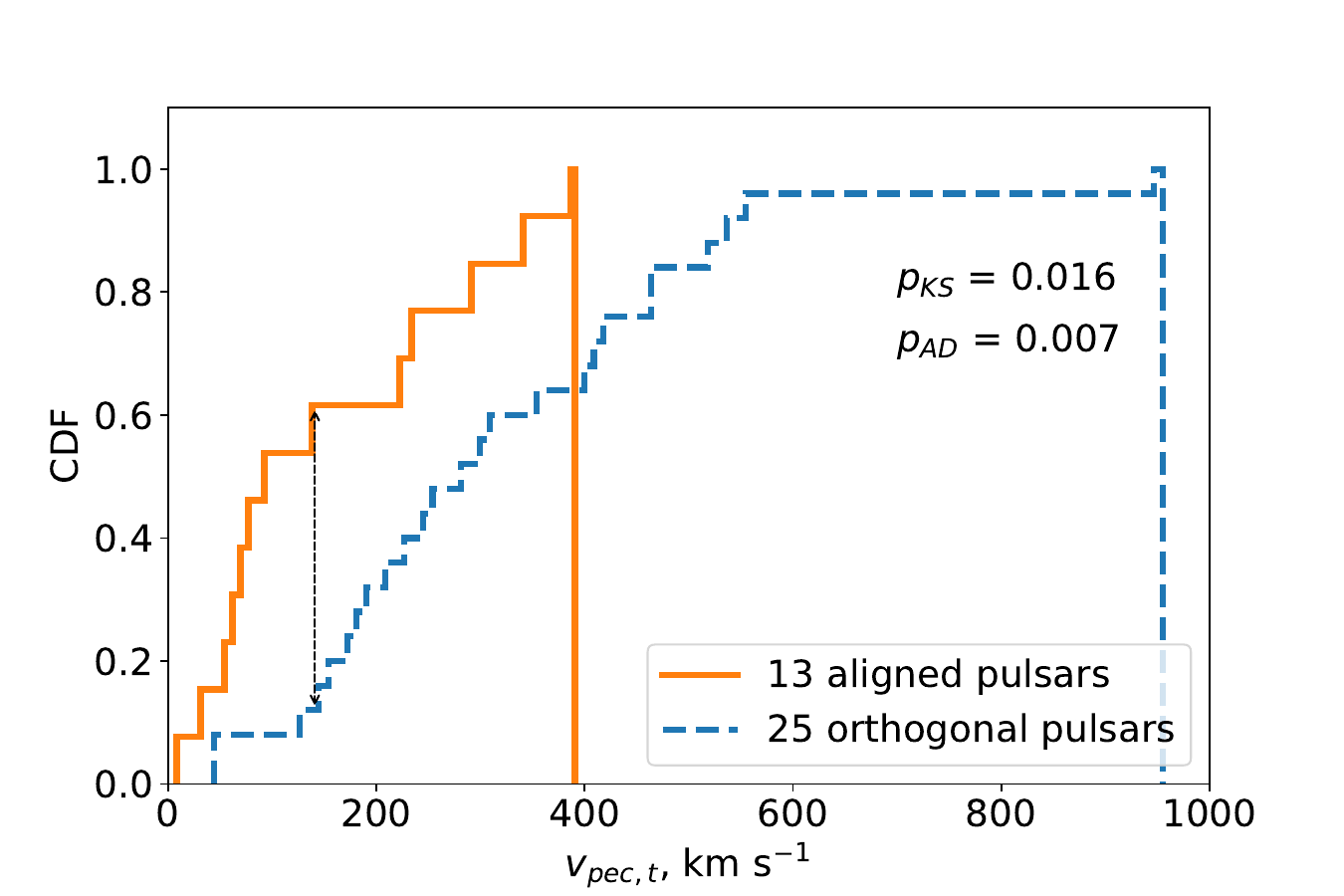}
\includegraphics[width=1\linewidth]{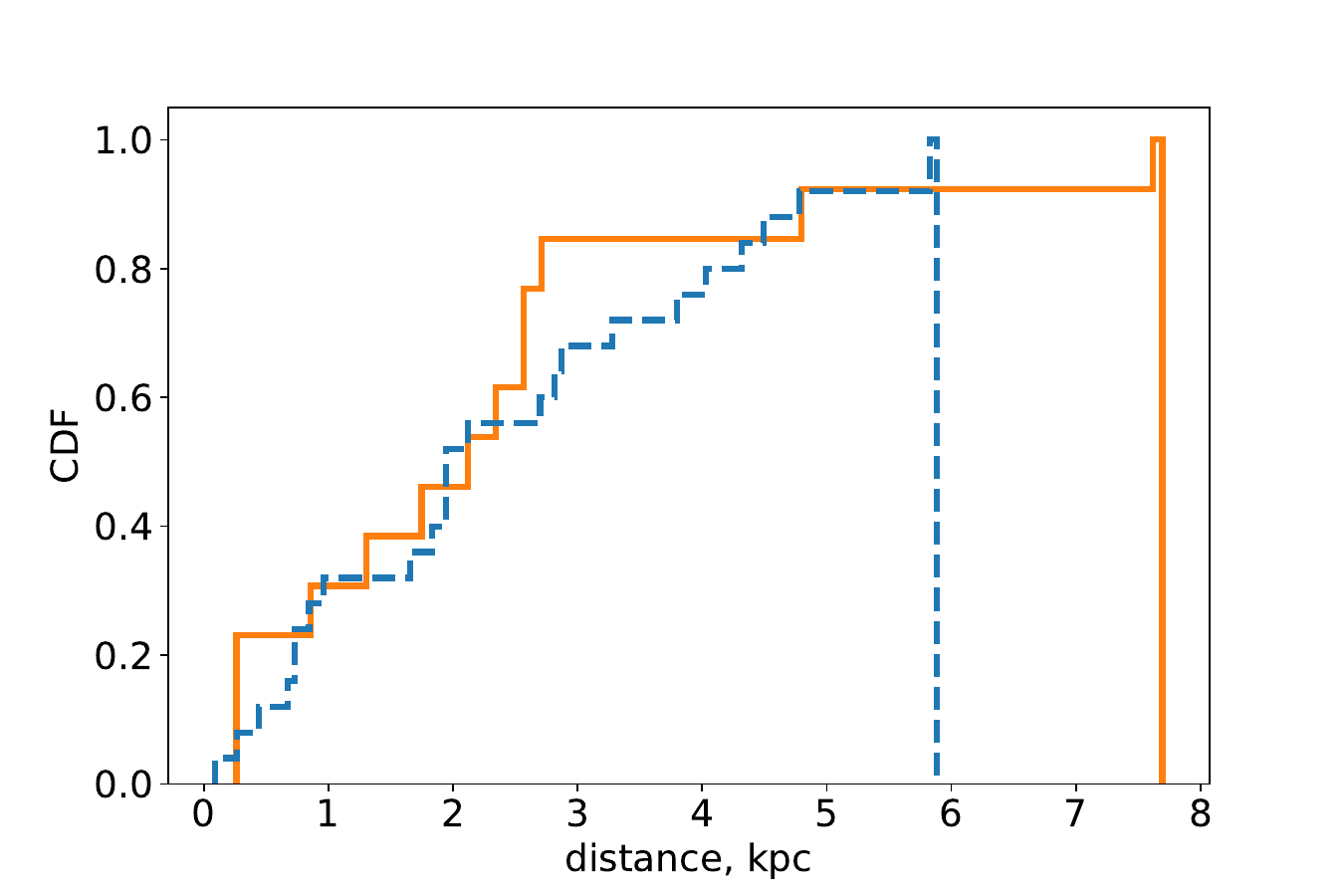}
\caption{{\bf Top plot}: Cumulative distributions of the absolute values of the transverse peculiar velocities for both types of pulsars. The solid orange line represents weakly oblique (``aligned'') rotators, while the dashed blue line represents strongly oblique (``orthogonal'') rotators. There are 13 aligned and 25 orthogonal pulsars were used, whose parameters are listed in Table~\ref{tab:pulsars} in \ref{app:pop_synth_results}. The two distributions differ: orthogonal rotators show systematically larger and more scattered velocities. The Kolmogorov–Smirnov two-sample test rejects the hypothesis that both subsets represent the same parent distribution at a $p_\mathrm{KS} = 0.016$ confidence level, and the Anderson–Darling test gives a $p_\mathrm{AD} = 0.007$. This divergence can be interpreted as an imprint of the pulsar spin-kick alignment. Magnetically aligned pulsars, in particular, tend to move along the line of sight, while orthogonal rotators move perpendicular to it. {\bf Bottom plot}: Cumulative distributions of estimated distances for both types of pulsars. Their similarity suggests that the difference detected in the top plot is not due to systematic differences in distance measurement.}
\label{fig:vt_observed}
\end{figure}

Therefore, in this work, we focus on the statistics of the projection of the pulsar's peculiar velocity onto the star's viewing plane.
\begin{equation}
    \mathbf{v}_{\mathrm pec,t} = \mathbf{v}_{\mathrm t} + \mathbf{v}_{\mathrm \odot,t} - \mathbf{v}_{\mathrm circ,t},
\end{equation}
where
\begin{equation}
    \mathbf{v}_{\mathrm \odot,t} = \mathbf{v}_{\mathrm \odot} - (\mathbf{v}_{\mathrm \odot} \cdot \mathbf n) \mathbf n
\end{equation}
is the projection of the Solar System's Barycenter galactic velocity. Unit vector $\mathbf n$ is directed towards the pulsar:
\begin{equation}
    \mathbf n = \left ( 
    \begin{array}{c}
    \cos l \cos b \\
    \sin l \cos b \\
    \sin b
    \end{array} 
    \right).
\end{equation}
In our calculations, we adopt an approximate value 
\begin{equation}
    \mathbf{v}_\odot = (10, 235, 7.5)~\mbox{km s}^{-1}, 
\end{equation}
which is well consistent with recent estimations \citep{2010MNRAS.403.1829S, 2024A&A...687A.272D}. Due to the uncertainties in estimating the distance of pulsars from their dispersion measures, we do not need to know this particular velocity with high accuracy. Furthermore, we also use a simplified, fully analytic model of the three-component gravitational potential $\phi(R, z)$, as offered by \cite{car87} and modified by \cite{kui89}, for the circular velocity $\mathbf{v}_{\mathrm circ,t}$. Calculating 
\begin{equation}
    \mathbf{v}_{\rm circ}(R) = \sqrt{R \cdot \nabla \phi(R, 0)}\left ( 
    \begin{array}{c}
    y\\
    -x + R_\odot \\
    0
    \end{array}
    \right ),
    \label{eq:v_circ}
\end{equation}
we assume it to be aligned with the galactic plane and depending only on the cylindrical radius $R = \sqrt{x^2 + y^2}$ and the galactocentric distance of the Sun $R_\odot = 8.5$ kpc, which is consistent with the adopted gravitational potential.

Figure~\ref{fig:vt_observed} (top) shows the cumulative distributions of the absolute values $|\mathbf{v}_{\mathrm{pec},t}|$ for both types of pulsars. The difference between the two distributions is clear: orthogonal rotators (shown in blue) have systematically larger and more scattered velocities. This is precisely what one would expect in the case of spin-kick alignment. The Kolmogorov–Smirnov two-sample test rejects the hypothesis that both subsets represent the same parent distribution at a $p = 0.016$ confidence level, and the Anderson–Darling test detects an even more significant difference: $p = 0.007$. However, this difference is not due to the systematics of the distance estimates, as the distributions of the distances for both groups are almost identical, as can be seen in the lower plot of Figure~\ref{fig:vt_observed}.

These distributions can be explained by assuming that aligned pulsars tend to move along the line of sight. Consequently, they exhibit low dispersion in their transverse velocities. This is consistent with the spin-kick alignment hypothesis. However, it is impossible to determine whether the kick is in the same direction as or opposite to the pulsar's spin axis. This is the first direct, kinematic, population-based evidence of pulsar spin-kick alignment.

We performed a detailed population synthesis of radio pulsars to provide a robust theoretical argument that this difference is indeed expected and detectable. The fundamental difficulty here lies in accurately accounting for the selection bias in estimates of transverse velocity. As the quantities in our subset have been measured using different instruments and techniques, our aim is merely to demonstrate the existence of the described distribution splitting effect and estimate its amplitude.

\section{Population synthesis}
\label{sect:popsynth}

For reasons of simplicity, in our work, we generally reproduces the well-established algorithm initially described by \cite{fgk06} (hereafter referred to as FK06) with modifications relevant to the analysis's aims. This section provides a detailed description of our model and computational setup.

\subsection{Pulsar initial Galactic positions}
\label{sect:newborn}
The coordinates and velocities of the synthetic pulsars were considered within a Galactocentric Cartesian right-handed reference frame, with the observer shifted from the centre for $R_{\odot} = 8.5\mbox{ kpc}$ and located within the equatorial plane. Following FK06, we simulated the positions of newborn pulsars along the four spiral arms established by \cite{gg76} and quantified by \cite{wai92}. All birth locations were then smoothed relative to the centroid of the spiral arms, as described in FK06.

However, the surface density of newborn pulsars was modelled slightly differently, with 
\begin{equation}
	p(R) = A \left (\frac{R}{R_\odot} \right )^4 \times \exp \left( -6.8 \cdot \frac{R}{R_\odot} \right ),
	\label{eq:yk04_birth}
\end{equation}
where $R$ is the galactocentric cylindrical radius and $A \approx 71.3$ is a normalisation constant \citep{yk04}. This distribution describes the localisation of young OB-type stars, which are believed to be neutron star progenitors. In FK06, another distribution was adopted for the same purpose. Specifically, the probability density function of {\it evolved} pulsars, $p(R)$, was used, which differs slightly from equation (\ref{eq:yk04_birth}) and was also found by \cite{yk04}. However, (\ref{eq:yk04_birth}) seems more physically motivated and, as demonstrated below, successfully reproduces observed pulsar statistics.

For pulsar birth places along the galactic $z$-axis (vertical), the double-sided exponential distribution with $\langle z_0 \rangle = 50$ pc has been used.

\subsection{Pulsar initial velocities}
The initial velocities of the synthetic pulsars were calculated as the vector sum of the progenitor's circular velocity (\ref{eq:v_circ}) and the isotropic kick velocity ${\mathbf v}_{\rm kick}$.  The absolute value of the latter was based on a double-sided exponential distribution
\begin{equation}
    p(v_{\mathrm 1D}) = \dfrac{1}{\langle v_{\mathrm 1D} \rangle} \exp \left (-\dfrac{|v_{\mathrm 1D}|}{\langle v_{\mathrm 1D} \rangle} \right)
    \label{eq:kick_distrib}
\end{equation}
with $\langle v_{\rm 1D} \rangle$ = 180 km s$^{-1}$ as was found by FK06. According to this, the average full three-dimensional velocity of newborn pulsars is $\langle v_{\rm kick, 3D} \rangle$ = 380 km s $^{-1}$. The kick velocity was considered to be independent of other pulsar parameters. However, numerical simulations of the NS formation process suggest a potential correlation between the initial spin period and the velocity of the stellar remnant \citep{2024ApJ...963...63B}. We will study this effect in further papers.

\subsection{Spin-kick alignment}
Two scenarios of the spin-kick relationship were investigated: isotropic and statistically aligned.

For each synthetic pulsar, a unit vector, $\mathbf w$, representing the orientation of its spin axis, was generated. In the case of an {\it isotropic kick}, the vector $\mathbf w$ was modelled independently of $\mathbf v_{\rm kick}$. In the case of an {\it aligned kick}, however, it was generated so that the angle between the kick velocity vector and the unit vector representing the orientation of the spin axis followed a zero-centred normal distribution with a standard deviation of $\sigma_{\mathrm kick} = 15^\circ$. This value is consistent with empirical estimations by \cite{nout12}. In particular, this value of $\sigma_{\mathrm kick}$ corresponds to a reasonably small value that allows the distribution of the angle between the pulsar velocity and spin axes projections onto the viewing plane to be reproduced (see also Section~\ref{sect:popsynth_results}).

\subsection{Pulsar physical parameters and their evolution}

\subsubsection*{Pulsar ages}
The ages of the synthetic pulsars were generated uniformly within the interval from 0 to 1.2 Gyr, corresponding to a constant galactic pulsar birthrate over the last 1.2 Gyr. The upper bound of this interval is determined by the longest pulsar lifetime within the adopted spin-down and deathline models (see below). The equation of motion for the synthetic pulsars, $\ddot {\mathbf r} = -\nabla \phi({\mathbf r})$, has been solved numerically until the prescribed age of the pulsar. During these calculations, we control the total energy of the pulsar (kinetic plus potential) and discard the result if it diverges by more than one per cent over the pulsar's lifetime.

\subsubsection*{Spin-down model}
\label{sect:spindown}
Unlike FK06, we adopt a more realistic pulsar spin-down model derived from MHD and PIC simulations \citep{spitkovsky06, phil14}. Thus, the evolution of the pulsar spin period, $P(t)$, is described by the equation
\begin{equation}
	P(t) \cdot \dfrac{dP(t)}{dt} = \xi B^2 \left[1 + 1.4\sin^2 \alpha(t) \right].
	\label{eq:spindown}
\end{equation}
Here, B is a constant surface (dipolar) magnetic field; $\alpha(t)$ is the variable angle between the pulsar magnetic and spin axes; and $\xi = 4\pi^2 R_{\mathrm NS}^6/Ic^3 = 1.75\times 10^{-39}$ sec/Gs$^2$. The neutron star radius $R_{\mathrm{NS}}$ and the moment of inertia $I$ in this equation were taken for a star with a mass of 1.5 $M_{\odot}$, assuming the WFF2 equation of state \citep{WFF}. Since we are interested in the evolution of classical pulsars with moderate magnetic fields of the order of $10^{12}$ G, we ignore their possible magnetic field decay. Indeed, due to the proximity of the expected Ohmic dissipation timescale to the typical pulsar lifetime, their magnetic fields do not change significantly during evolution \citep[e.g][]{ip21, agu08}.

The evolution of the magnetic angle, $\alpha(t)$, was modelled consistently with equation (\ref{eq:spindown}) as
\begin{equation}
	P(t)^2 \cdot \dfrac{d\alpha(t)}{dt} = - \xi B^2 \sin\alpha(t) \cdot \cos\alpha(t).
	\label{eq:obliquity_evolution}
\end{equation}
The distribution of the initial pulsar periods $P_0 = P(0)$, was assumed to be normal with an average $\langle P_0 \rangle$ and a standard
deviation $\sigma[P_0]$. These two parameters were estimated during the optimisation procedure. The same is true for the distribution of magnetic fields (described via $\langle \log B \rangle$ and $\sigma[\log B]$, respectively).

At the same time, initial values for the magnetic angle, $\alpha_0 \in 0..90^{\circ}$, were taken from the isotropic distribution so that 
\begin{equation}
	p(\alpha_0) = \dfrac{1}{2} \sin\alpha_0.
	\label{eq:initial_alpha}
\end{equation}

\subsubsection*{Radio luminosity}
The pulsar pseudo-luminosity\footnote{Conventional pseudo-luminosity enables the observed radio flux of a pulsar to be predicted statistically rather than reflecting its actual radio luminosity.} at a frequency of 1.4 GHz was modelled in a manner similar to that described by \cite{gullon14}:
\begin{equation}
	L_\mathrm{1.4GHz} = L_0 \cdot 10^{L_\mathrm{corr}} \cdot \left( P^{-3} \dot P 
	\right)^{1/2},
	\label{eq:luminosity}
\end{equation}
where the pulsar period and its derivative are expressed in seconds and seconds per second, respectively. According to FK06, the correction parameter $L_\mathrm{corr}$ was assumed to be normally distributed with a zero mean and a standard deviation $\sigma[L_\mathrm{corr}] = 0.8$. The constant $L_0$ (mJy kpc$^2$) is the final free parameter in our simulations and was assumed to remain constant for all pulsars throughout a run. Synthetic radio flux was then defined as
\begin{equation}
	F_\mathrm{1.4GHz} = \dfrac{L_\mathrm{1.4GHz}}{d^2},
	\label{eq:flux}
\end{equation}
where $d$ is the modelled distance to the pulsar from the Solar System Barycentre in kiloparsecs.

\subsubsection*{Death line}
The pulsar death line was adopted in the form
\begin{equation}
	\dot P_{\rm crit} = (2.82 \cdot 10^{-17}\mbox{ sec}^2) P^3
	\label{eq:death_line}
\end{equation}
which was initially established by \cite{raw86} from the observations of a long-period pulsar. This death line is equivalent to the equation used
by FK06 ($B/P^2 < 0.17 \times 10^{12}$ Gs/sec$^2$) if one assumes $B = 3.2\cdot 10^{19} \sqrt{P \dot P}$ Gs -- the standard estimation of the pulsar magnetic field. As a purely empirical filter, equation (\ref{eq:death_line}) remains close to the classical theoretical prediction $\dot P_{\rm crit} \propto P^{2.25}$ by \cite{rs75}.

\subsection{Pulsar observational selection}

\subsubsection*{Viewing geometry}
\label{sect:w10}

We assume that synthetic pulsars emit radiation within the two identical, symmetric pencil-shaped beams, which are directed along the magnetic axis. The radius $\rho$ of each beam was calculated according to Rankin's formula for the opening angle of pulsar outer conal emission
\begin{equation}
	\rho = 5.7^\circ \mbox{ }P^{-1/2},
	\label{eq:beam_radius}
\end{equation} 
where $P$ is in seconds \citep{rankin93b}.

The width of the unscattered pulse $w_{10}$ (at a tenth of the maximum) has been calculated for every synthetic pulsar. Geometrically, it follows the equation
\begin{equation} 
	\cos \left ( \frac{w_{10}}{2} \right) = C(\alpha, \rho, \theta) = \frac{\cos\rho - \cos\alpha\cos\theta}{\sin\alpha \sin\theta},
	\label{eq:cosW10}
\end{equation}
where $\theta \in 0..180^\circ$ is the observer's obliquity with respect to the NS spin axis: $\cos\theta = -(\mathbf w \cdot \mathbf n)$. 
The pulse width (\ref{eq:cosW10}) has been calculated for both  pulsar beams so that $C_{\rm n} = C(\alpha,\rho,\theta)$ for ``north'' beam and $C_{\rm s} = C(180^\circ - \alpha,\rho,\theta)$ for the ``south'' one respectively.

The decision as to whether the simulated pulsar is ``directed'' towards the observer at time $t$ was made as follows:
\begin{itemize}
	\item If $|C_{\rm n}| < 1$ then pulsar is detectable and 
	$w_{10} = 2\arccos(C_{\rm n})$ independently on the value $C_{\rm s}$;
	\item If $|C_{\rm n}| > 1$ and $|C_{\rm s}| < 1$ then pulsar is detectable and $w_{10} = 2\arccos(C_{\rm s})$;
	\item If $|C_{\rm n}| > 1$ and $|C_{\rm s}| > 1$ then pulsar cannot be detected.
\end{itemize}

\subsubsection*{Detection threshold and control subset}
\label{sect:sky}
We have almost replicated the algorithm described in FK06 in order to model telescope sensitivity. Specifically, our aim was to reproduce pulsars that could be detected by the Parkes and Swinburne Multibeam Surveys \citep{man01, edw01}. Within these surveys, 1057 isolated, rotation-powered pulsars were detected at a central frequency of 1.4 GHz and found in the ATNF database. However, note that $W_{10}$ values are known only for 381 of these pulsars, and full proper motion ($v_t$) is known for only 106 of them.

In order to estimate survey sensitivity, the background brightness temperature, $T_\mathrm{sky}$, is required. We calculated it using a slightly different method to that employed by FK06. Specifically, while FK06 used an electronic version of the $T_\mathrm{sky}$ maps obtained by \cite{has81} at a central frequency of 408 MHz, in our work we adopted the analytical approximation of the same maps found by \cite{nar87} in the form
\begin{equation}
	T_{\rm sky}({\rm @408 MHz}) = 25 + \frac{275}{[1 + (l/42)^2]\cdot [1 + (b/3)^2]}\mbox{ K},
\end{equation}
where $l$ and $b$ are galactic longitude and latitude, respectively, taken in degrees. We then scaled $T_\mathrm{sky}$ to 1.4 GHz according to a power law with a spectral slope of -2.8 \citep{law87}.  The brightness temperature at the sky coordinates of a synthetic pulsar, as well as the dispersion measure, DM, towards its position\footnote{Which was calculated using the galactic electron density model by \cite{ne2001} to be consistent with the initial FK06 setup}, were then used to estimate interstellar distortion of the signal and the survey detection threshold, $F_{1.4, min} \propto T_{\rm sky} \sqrt{DM}$. The observed pulse width, $W_{10}$ also took into account interstellar dispersion and telescope parameters.

\section{Results}
\label{sect:results_main}

\subsection{Pulsar population properties}
\label{sect:popsynth_results}
We considered a discrete grid in the space of the free parameters of our model. There are five of these parameters: two describing the distribution of the initial period ($\langle P_0 \rangle$ and $\sigma[P_0]$), two describing magnetic fields ($\langle \log B \rangle$ and $\sigma[\log B]$), and one describing the luminosity-period relationship ($L_0$). The grid steps were set to 0.05 dex for the initial spin period and magnetic field distribution parameters, and to 0.125 dex for the luminosity normalisation constant $L_0$. The values corresponding to the greatest similarity between the synthetic and observed distributions are listed in Table~\ref{tab:synthesis_results}. The number of potentially detectable galactic pulsars and their estimated birth rate are also provided.

\begin{table}
\begin{threeparttable}
\caption{Population synthesis best parameters}
\label{tab:synthesis_results}
\begin{tabular}{ll}
\toprule
\headrow Parameter & Value  \\
\midrule
Average initial period $\langle P_0 \rangle$ & 0.3 sec \\ 
\midrule
Initial period dispersion $\sigma[P_0]$ & 0.2 sec \\ 
\midrule
Average magnetic field $\langle \log \left (B/\mathrm{Gs} \right) \rangle$ & 12.45 \\ 
\midrule
Magnetic field dispersion $\sigma [\log \left (B/\mathrm{Gs} \right)]$ & 0.6 \\ 
\midrule
Luminosity constant $L_0$ & 4.27 $\cdot$ 10$^6$ mJy kpc$^2$ \\
\midrule
\midrule
\headrow Derived parameters: isotropic kick & \\
\midrule
Pulsar average lifetime & 2.4 Myr\\
\midrule
Potentially detectable pulsars in Galaxy & 62,000 \\
\midrule
Pulsar average birthrate & 2.9 $\pm$ 0.1 century$^{-1}$\\
\midrule
\headrow Derived parameters: correlated kick & \\
\midrule
Pulsar average lifetime & 3.0 Myr\\
\midrule
Potentially detectable pulsars in Galaxy & 74,000 \\
\midrule
Pulsar average birthrate & 2.8 $\pm$ 0.1 century$^{-1}$\\
\bottomrule
\end{tabular}
%\begin{tablenotes}[hang]
%\item[b]Another table note
%\end{tablenotes}
\end{threeparttable}
\end{table}

\begin{figure}[hbt!]
\centering
\includegraphics[width=1\linewidth]{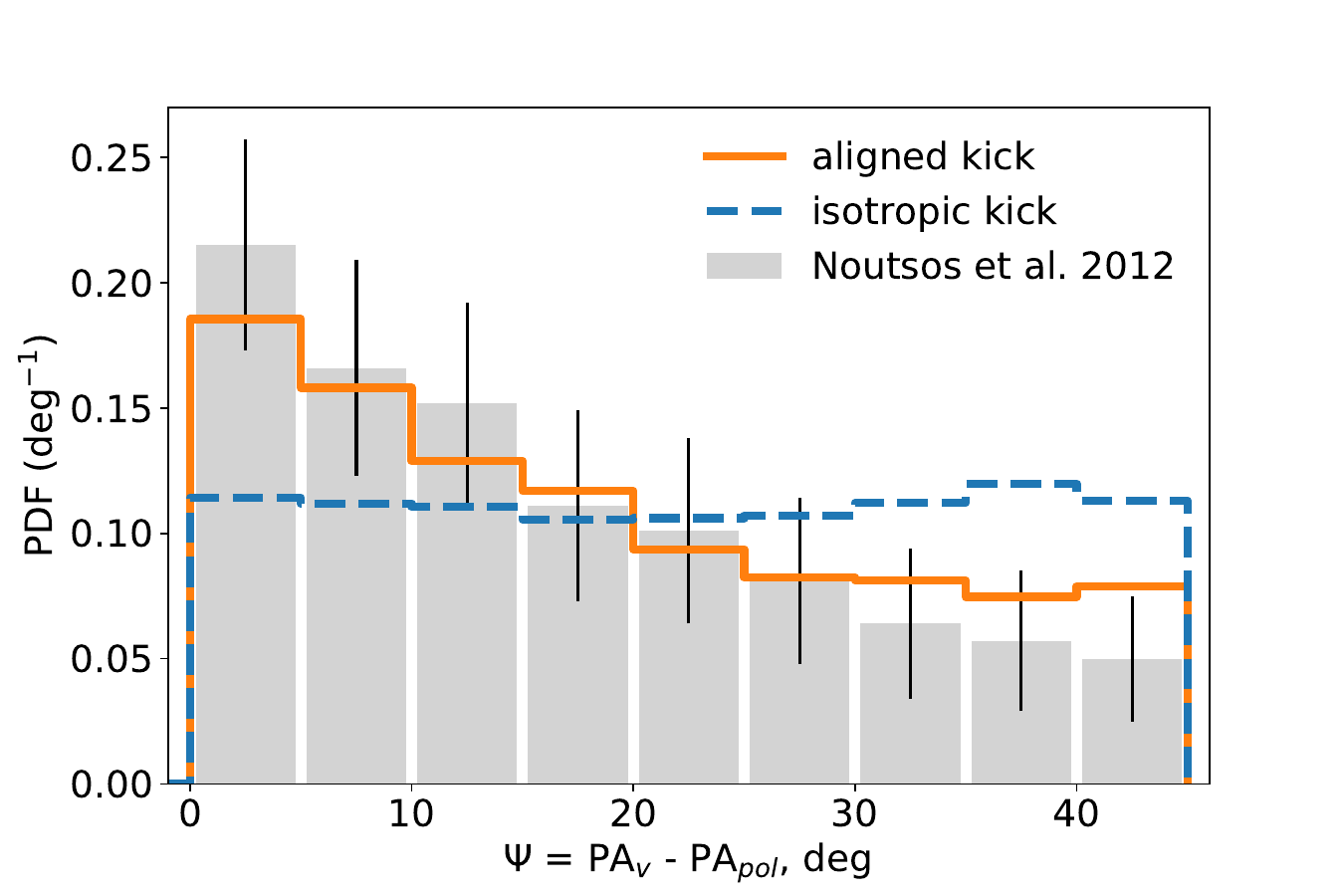}
\caption{
Distributions of the angle $\Psi = $PA$_\mathrm{v}$ - PA$_\mathrm{pol}$, representing the divergence between the proper motion vector and the projection of the pulsar spin axis onto the viewing plane. The grey bars show the observed distribution, taken from \cite{nout12}, based on 54 pulsars. Values are normalised from 0 to 45 degrees due to degeneracy of the Rotating Vector Model regarding the $X-$ and $O-$ polarisation modes in pulsar emission. Angle $\Psi$ is the only parameter sensitive to the spin-velocity correlation. An isotropic kick scenario (shown by the blue dashed line) results in a nearly uniform distribution of $\Psi$, which is inconsistent with observations. Conversely, a spin-aligned kick scenario (the solid orange line) produces a more suitable, non-uniform distribution.}
\label{fig:psi_compare}
\end{figure}

These parameters are consistent with those obtained in FK06 and in a more recent analysis by \cite{Igoshev22}. Interestingly, our model predicts short pulsar lifetimes of 2.5–3 Myr on average, although much older pulsars are undoubtedly possible. This estimation is reasonable, but could be affected by some incompleteness in our understanding of pulsar beam width, the luminosity model, telescope selection effects and the spin-down law (see the discussion below). Within this time interval, however, one could expect a pulsar's peculiar velocity to still reflect its natal kick velocity. This explains the distribution split observed in Figure~\ref{fig:vt_observed}.

The corresponding synthetic distributions of pulsar observables are shown in Figure~\ref{fig:synthesis_results} of the \ref{app:pop_synth_results}. 
The solid orange and dashed blue lines represent the aligned and isotropic kick scenarios, respectively. The grey-filled histograms show the observed distributions for 1057 control objects. Following \cite{fgk06}, we plot the classical magnetic field estimation instead of the spin period derivative:
\begin{equation}
    B_{\rm md} = 3.2\cdot 10^{19} \sqrt{P \dot P}\mbox{ Gs},
    \label{eq:Bmd}
\end{equation}
where $P$ and $\dot P$ are taken in seconds and sec/sec respectively\footnote{Note,
that coefficient of $3.2\cdot 10^{19}$ Gs in equation (\ref{eq:Bmd}) is formally calculated for a 10 km NS with $I = 10^{45}$ g cm$^2$ assuming a simple magnetodipolar spin-down model and $\alpha = 90^{\circ}$. Therefore, $B_{\rm md}$ is a quantity that substitutes for $\dot P$ rather than being a robust estimation of a pulsar's magnetic field. See also \cite{bab17} for additional discussion. }. We conclude that the obtained synthetic distributions are close to the observed ones.

\begin{figure*}
\centering
\includegraphics[width=1\linewidth]{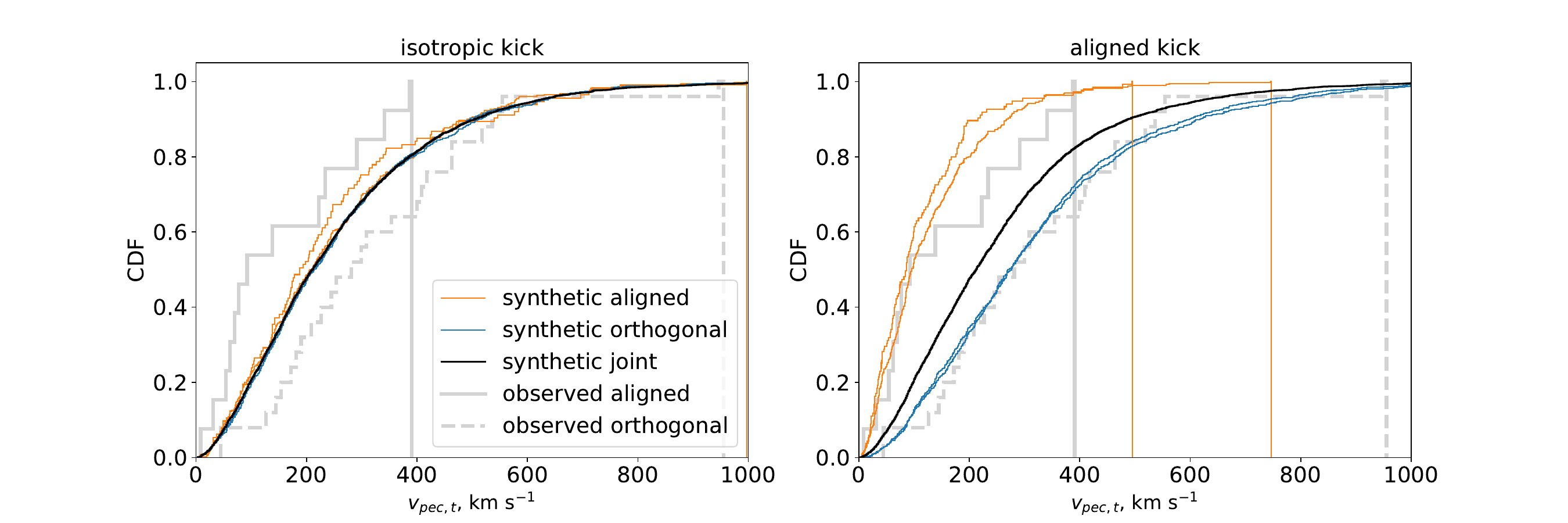}
\caption{These plots illustrate the paper's primary theoretical outcome. The distributions of the modelled (synthetic) peculiar velocities, $v_{\mathrm{pec}} = v - v_{\mathrm{LSR}}$, are shown after projection onto the viewing plane. The left plot contains results for an isotropic kick velocity, and the right plot contains results for a spin-aligned velocity. The solid black lines show the overall distribution of pulsar velocities. However, observational selection effects are taken into account in the same way as for the distributions shown in Figures ~\ref{fig:synthesis_results} and ~\ref{fig:synth_derivs}. The thin orange and blue lines on both plots represent the velocity distributions of pulsars that are nearly aligned (with $\alpha$ < 25 and 10 degrees) and nearly orthogonal (with $\alpha$ > 65 and 80 degrees), respectively. A clear effect seen in the simulated data is the splitting of the velocity distribution for these two types of pulsar in the case of spin-kick alignment. Specifically, weakly oblique pulsars have systematically smaller and less dispersed observed transverse velocities than orthogonal ones. At the same time, isotropy of the kick destroys this effect. The observed distributions from Figure~\ref{fig:vt_observed} are also shown as light grey lines on the plots. Notably, the observed and synthetic distributions are very close to each other in the case of spin-kick alignment, even though this was not the aim of the calculations. In general, this plot shows that spin-kick alignment manifests as different transverse velocities for aligned and orthogonal pulsars in a synthetic pulsar galaxy with realistic underlying properties. Furthermore, the strength of this difference is very close to that observed.}
\label{fig:main_result}
\end{figure*}

Additionally, Figure~\ref{fig:synth_derivs} in \ref{app:pop_synth_results} shows the distributions of further pulsar parameters that are important for analysing their kinematics and viewing geometry. These are the magnetic angle $\alpha$, the pulse width at 10 per cent of the maximum ($W_{10}$), the proper motion components $\mu_l \cos b$ and $\mu_b$, the transverse velocity $v_t$ and the distance $d$.

These parameters do not reproduce the observed distributions very well. However, this is not unexpected, as we ignored any selection bias that could affect estimations of pulsar proper motions, distances and, therefore, transverse velocities. In our model, synthetic pulsars have systematically larger observed velocities of approximately 150 km s$^{-1}$. We assume that at least two types of selection effect can account for this difference. The first is systematic errors in estimating the velocities of slow pulsars, which lead to the false detection of small $v_t$. The second is the small number of pulsars used to verify the kick velocity distribution (\ref{eq:kick_distrib}). Thus, in FK06, only the sample of 34 pulsars observed by \cite{Brisken02} and \cite{Brisken03} was used. Ultimately, this slight inconsistency cannot affect our results. However, we can conclude that none of these parameters are sensitive to the specific spin-kick alignment scenario.

The only parameter that exhibits such sensitivity is the angle $\Psi$, which is the angle between the projections of the pulsar's spin and velocity vectors. Its properties have been investigated in detail previously  \citep{Johnston05, nout12, nout13}. It exhibits a non-uniform distribution, which is the primary statistical evidence for the pulsar spin-kick correlation. The orientation of the spin axis in the sky is associated with the position angle, $\mathrm{PA}_\mathrm{pol}$, of the linear polarisation of the pulsar's emission. In particular, $\mathrm{PA}_\mathrm{pol}$ is the position angle at the maximum derivative of the $\mathrm{PA}$ swing. However, observations cannot distinguish between clockwise and counterclockwise rotating pulsars. This leads to an uncertainty of 90 degrees in $\Psi$. Additionally, we cannot differentiate between the X- and O-modes of pulsar polarisation, which exacerbates this uncertainty. As a result, the observable $\Psi$ is the smallest angle between the direction of the proper motion $\mathrm{PA}_\mathrm{v}$ and one of the two axes, one of which is given by the position angle $\mathrm{PA}_\mathrm{pol}$ and the other by $\mathrm{PA}_\mathrm{pol} + 90^\circ$

Figure~\ref{fig:psi_compare} shows synthetic distributions of $\Psi$ obtained in population synthesis and the observed distribution from \cite{nout12}. We conclude that an isotropic kick cannot reproduce the observations, whereas a spin-aligned kick can. This result is consistent with that obtained by \cite{nout12}, but is based on an extensive, detailed population synthesis.

\subsection{Velocities of aligned and orthogonal rotators}
However, the main parameter of interest in the current work is the peculiar transverse velocity of the pulsars. Figure~\ref{fig:main_result} shows the modelled distributions of this quantity, which are similar to those in Figure~\ref{fig:vt_observed}. The solid black lines show the resultant distributions involving every synthetic pulsar, which look almost identical in both spin-kick relationship scenarios. However, when weakly and strongly oblique pulsars are considered separately, a high spin-kick alignment produces a clear distribution split. This can be seen in the right-hand plot in Figure~\ref{fig:main_result}. A couple of distributions corresponding to nearly aligned synthetic pulsars with $\alpha$ less than 25 or 10 degrees are shown by thin orange lines. Similar distributions for almost orthogonal objects (with $\alpha$ greater than 65 and 80°) are shown by thin blue lines. These two sets of distributions are separated from each other, as well as from the joint distribution. In particular, aligned pulsars show systematically smaller and less scattered transverse velocities. This is precisely expected in the case of strong spin-kick alignment, which is observed in real pulsars. The real distributions from Figure~\ref{fig:vt_observed} are also shown on these plots by light grey lines.

It is notable that the modelled distributions on the right-hand plot agree well with the observed ones. While this coincidence was not the objective of the simulations, it likely reflects the adequacy and realism of the constructed population synthesis model.

Therefore, we conclude that the population synthesis performed supports the idea that the observed difference in the velocities of aligned and orthogonal pulsars is due to spin-kick correlation. Furthermore, kick velocities tend to align with the spin axis's direction.

\section{Discussion}
\label{sect:discuss}

\subsection{Realistic rotational evolution of radiopulsars}

Although the equations of pulsar spin-down (\ref{eq:spindown}) and (\ref{eq:obliquity_evolution}) are confidently constrained from an accurate numerical study, they still remain purely theoretical results. The long-term rotational evolution of real pulsars appears to be more complex\footnote{Hereafter, we will not discuss pulsar timing noise and glitches as these only affect the rotational evolution only at very short timescales of months and years. Interested reader are reffered to \cite{2024Galax..12....7A, zhou_glitches_review22, hobbs10} and references therein.}.  Thus, the so-called braking indices $n_\mathrm{br} = 2 - P\ddot P/\dot P^2$, are expected to lie within the range $3..3.25$ during a pulsar lifetime, assuming a constant magnetic field \citep{phil14}. However, for actual classic pulsars, the values of $n_\mathrm{br}$ spread in $\sim -10^5..10^5$, being even negative in half of the cases \citep{hobbs04, bbk07}. This challenges the adopted spin-down model. Assuming that the pulsar evolves with a huge, constant braking index, one gets that it have to either stop spinning down on timescales $\sim \tau_\mathrm{ch}/n_\mathrm{br}$ for positive $n_\mathrm{br}$, or completely lose its rotational energy ($P = 0$) after $\sim \tau_{ch}$ for $n_\mathrm{br} < 0$. This is the problem of anomalous braking indices. Many solutions have been proposed for it so far, but none have been widely accepted.

On the other hand, pulsar population studies show that smooth spin-down laws (\ref{eq:spindown})-(\ref{eq:obliquity_evolution}) really help to satisfactorily reproduce pulsar statistics. The population synthesis undertaken in this study provides further evidence for this. This contradiction can be resolved by assuming that braking indices oscillate with a large amplitude on timescales much shorter than the pulsar lifetime. Earlier, we have estimated this timescale as $\sim 10^3..10^4$ years \citep{bbk12}. The effect of such oscillations is averaged out during long-term evolution, which the current work is focused on. Therefore, we believe that the model (\ref{eq:spindown})-(\ref{eq:obliquity_evolution}) is still reliable and produces physically reasonable results.

\subsection{Orthogonal natal kick}
In our analysis, we have focused on a spin-aligned natal kick of pulsars. An alternative scenario -- when kick velocity is nearly orthogonal to the spin axis -- was not considered in detail. Nevertheless, we have performed simple Monte Carlo calculations for this case to ensure completeness. Thus, we treated neutron stars as rotators with magnetic angles that are strictly equal to 0 or 90 degrees. In other words, two extreme cases were considered in terms of the magnetic alignment. Of course, magnetic angle $\alpha = 0$ suppresses any pulsar effect. But, in our calculations, we are interested in statistics of the spin-velocity angles, and want to probe the strongest possible effect. Thus, adopting $\alpha = 0$ is still makes sense in this particular case.

\begin{figure}[hbt!]
\centering
\includegraphics[width=1\linewidth]{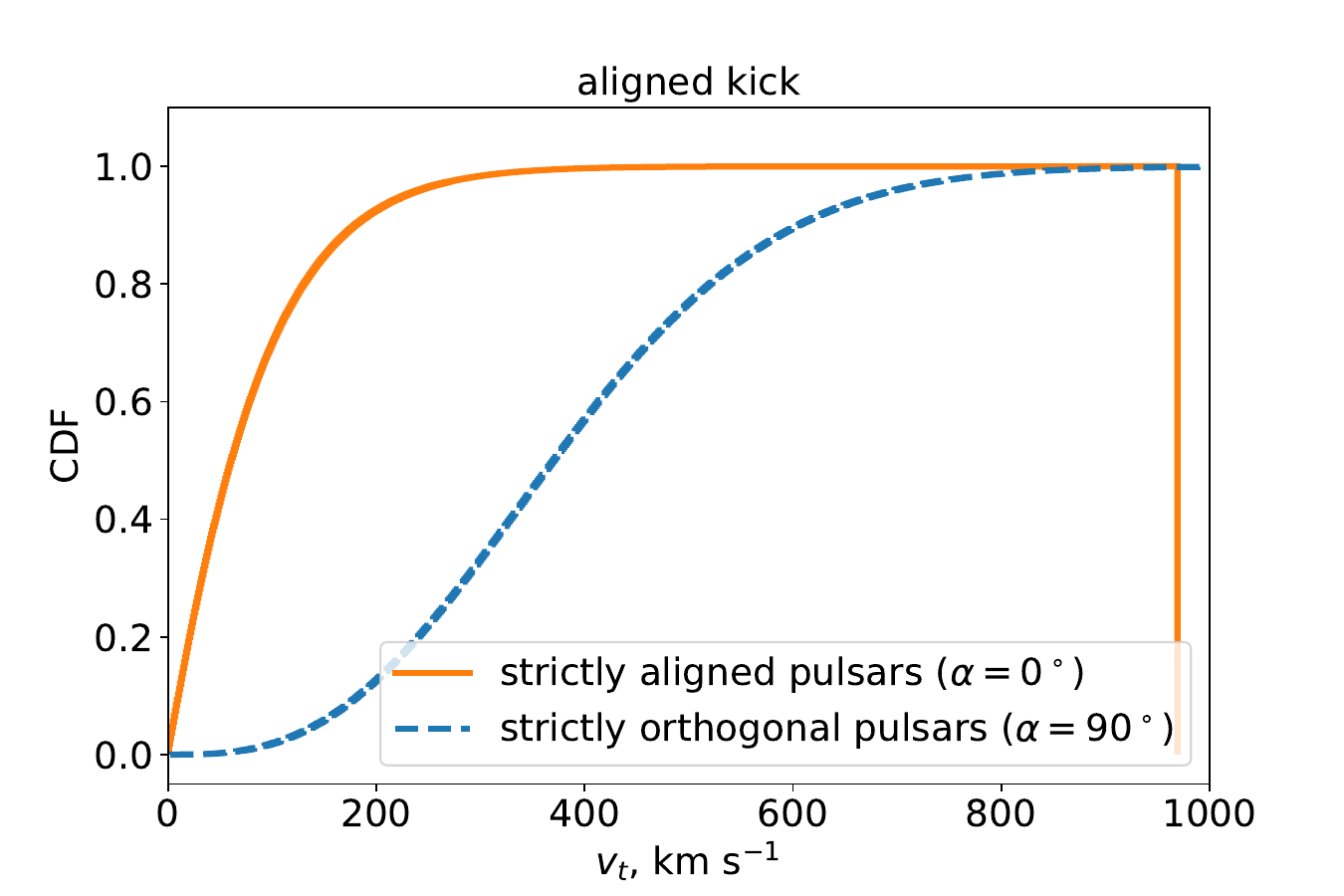}
\includegraphics[width=1\linewidth]{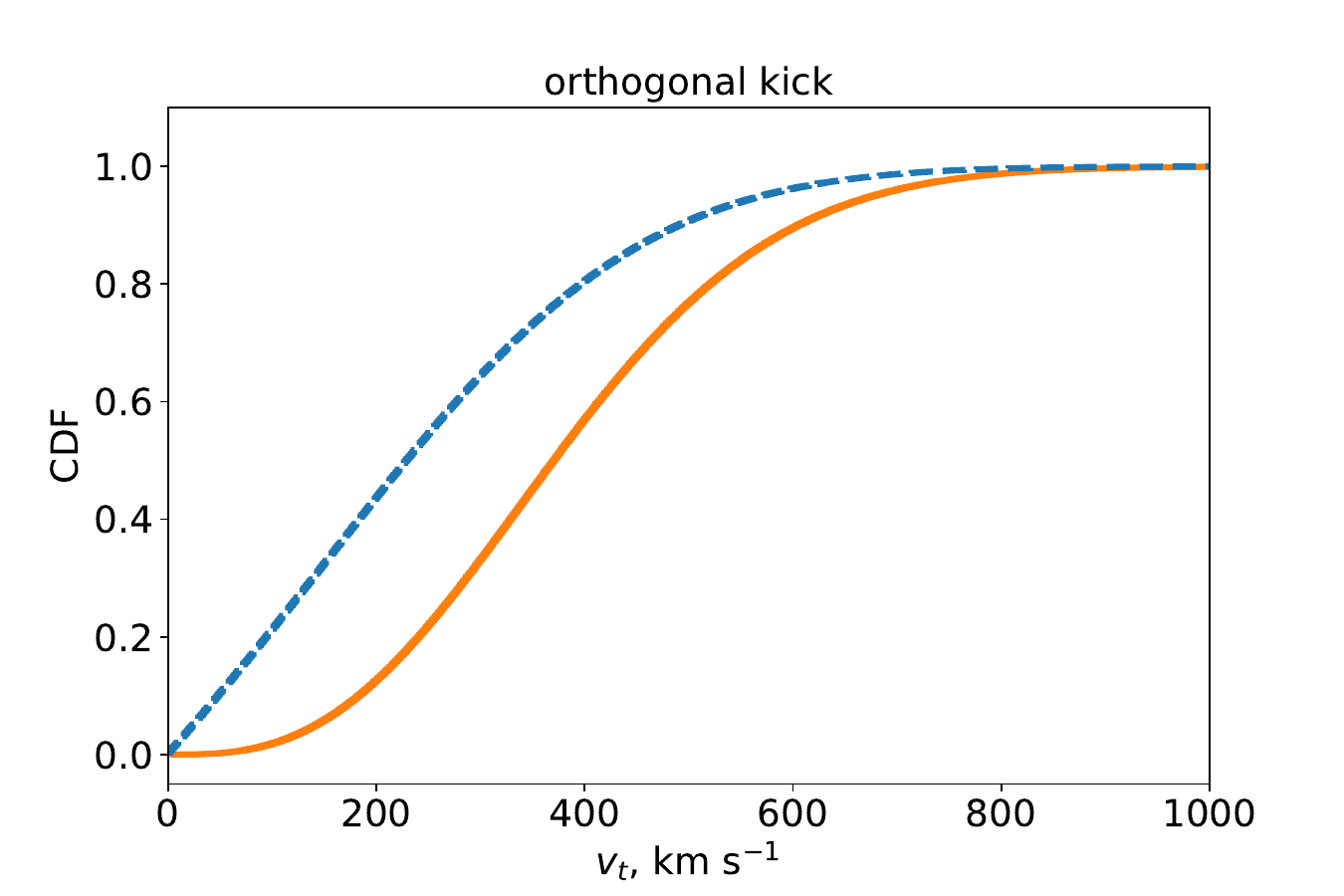}
\caption{These are the results of a simple Monte Carlo calculation of the distribution of transverse velocities of pulsars. The top plot shows the distribution of spin-aligned pulsar velocities for magnetically aligned and orthogonal pulsars. The difference between these distributions is consistent with that observed in population synthesis. In contrast, the bottom plot shows a similar distribution for a kick perpendicular to the spin axis. The split between the two distributions still occurs in this case, but the sign is different. Magnetically orthogonal pulsars, in particular, have qualitatively smaller and less dispersed transverse velocities. This supports the conclusion that Figure~\ref{fig:vt_observed} reflects spin-kick alignment, rather than orthogonality. See the text for details.}
\label{fig:test_orth}
\end{figure}

To generate the new synthetic subset of neutron stars, we modelled the velocity vector of each star in two ways. The first method involved setting the angle to the spin axis as normally distributed with a dispersion of 15 degrees. The second method is the same, but for the angle between the velocity vector and the star's equatorial plane. The azimuthal orientation of the velocity was chosen uniformly in the 0..360 degrees range. The absolute values of the velocity were taken from a Maxwellian distribution with an average of 250 km s$^{-1}$. By modelling $10^7$ such synthetic pulsars, we calculated the theoretical distributions of transverse velocities, i.e. the projections onto a common plane. These distributions are shown in Figure~\ref{fig:test_orth}. The case of a spin-aligned kick is shown on the top plot of this figure. The top plot of this figure shows the case of a spin-aligned kick. The difference in the transverse velocities of strictly aligned and orthogonal pulsars is exactly the same as that observed (see Figure~\ref{fig:vt_observed}) and obtained in the population synthesis (see Figure~\ref{fig:main_result}).

Conversely, a spin-orthogonal kick produces the inverse difference in these distributions. In particular, magnetically orthogonal pulsars show smaller and less dispersed velocities than magnetically aligned ones. Therefore, we conclude that the observed diversity in the velocities of the two types of pulsar indicates alignment, rather than orthogonality of their spin and 3D velocity direction. This is the first time that such evidence has been obtained for a subset of pulsars, but not for individual ones, independently of their polarisation.

\section{Conclusions}
\label{sect:conclude}
This work has shown the power of the pulsar velocities analysis method in revealing the spin-kick alignment phenomenon. This effect was initially established after analysing 54 pulsars \citep{nout12} more than a decade ago. However, recent massive estimations of pulsar polarisation parameters, including magnetic angles for more than 400 southern \citep{Meerkat_RVM} and almost 200 northern \citep{FAST_polarization23} pulsars, open a new possibility to refine underlying distributions and provide a more detailed understanding of this phenomenon. At the same time, it is evident that transverse velocities of relatively young orthogonal pulsars are the best estimations of their actual 3D velocities, which can be constrained for the first time. This fact could also be used for further analysis, which will be presented in future works.

\begin{acknowledgement}
The authors are thankful to Sergey Popov and Andrei Igoshev for their useful suggestions.
\end{acknowledgement}

\paragraph{Funding Statement}
The work was carried out within the framework of the state
assignment of the Special Astrophysical Observatory of the
Russian Academy of Sciences, approved by the Ministry of
Science and Higher Education of the Russian Federation

\paragraph{Competing Interests}
None.

\paragraph{Data Availability Statement}

All the data related to the described analysis can be provided upon request.

\printendnotes

\printbibliography

@ARTICLE{ne2001,
   author = {{Cordes}, J.~M. and {Lazio}, T.~J.~W.},
    title = "{NE2001.I. A New Model for the Galactic Distribution of Free Electrons and its Fluctuations}",
  journal = {ArXiv Astrophysics e-prints},
   archiveprefix = {astro-ph/},
   eprint = {0207156},
     year = 2002,
    month = jul,
   adsurl = {http://adsabs.harvard.edu/abs/2002astro.ph..7156C}
}

@ARTICLE{2009ApJ...698..250C,
       author = {{Chatterjee}, S. and {Brisken}, W.~F. and {Vlemmings}, W.~H.~T. and {Goss}, W.~M. and {Lazio}, T.~J.~W. and {Cordes}, J.~M. and {Thorsett}, S.~E. and {Fomalont}, E.~B. and {Lyne}, A.~G. and {Kramer}, M.},
        title = "{Precision Astrometry with the Very Long Baseline Array: Parallaxes and Proper Motions for 14 Pulsars}",
      journal = {Astrophys. J},
         year = 2009,
        month = jun,
       volume = {698},
       number = {1},
        pages = {250-265},
          doi = {10.1088/0004-637X/698/1/250},
archivePrefix = {arXiv},
       eprint = {0901.1436},
 primaryClass = {astro-ph.SR},
       adsurl = {https://ui.adsabs.harvard.edu/abs/2009ApJ...698..250C}
}

@article{agu08,
 adsurl               = {http://adsabs.harvard.edu/abs/2008A%26A...486..255A},
 archiveprefix        = {arXiv},
 author               = {{Aguilera}, D.~N. and {Pons}, J.~A. and {Miralles}, J.~A.},
 doi                  = {10.1051/0004-6361:20078786},
 eprint               = {0710.0854},
 journal              = {Astron. and Astrophys.},
 month                = {jul},
 pages                = {255-271},
 title                = {{2D Cooling of magnetized neutron stars}},
 volume               = {486},
 year                 = {2008},
 }

@ARTICLE{1998ApJ...505..315C,
       author = {{Cordes}, J.~M. and {Chernoff}, David F.},
        title = "{Neutron Star Population Dynamics. II. Three-dimensional Space Velocities of Young Pulsars}",
      journal = {Astrophys. J},
         year = 1998,
        month = sep,
       volume = {505},
       number = {1},
        pages = {315-338},
          doi = {10.1086/306138},
archivePrefix = {arXiv},
       eprint = {astro-ph/9707308},
 primaryClass = {astro-ph},
       adsurl = {https://ui.adsabs.harvard.edu/abs/1998ApJ...505..315C}
}

@ARTICLE{2022MNRAS.517.3938C,
       author = {{Coleman}, Matthew S.~B. and {Burrows}, Adam},
        title = "{Kicks and induced spins of neutron stars at birth}",
      journal = {MNRAS},
         year = 2022,
        month = dec,
       volume = {517},
       number = {3},
        pages = {3938-3961},
          doi = {10.1093/mnras/stac2573},
archivePrefix = {arXiv},
       eprint = {2209.02711},
 primaryClass = {astro-ph.HE},
       adsurl = {https://ui.adsabs.harvard.edu/abs/2022MNRAS.517.3938C}
}

@ARTICLE{2002ApJ...568..289A,
       author = {{Arzoumanian}, Z. and {Chernoff}, D.~F. and {Cordes}, J.~M.},
        title = "{The Velocity Distribution of Isolated Radio Pulsars}",
      journal = {Astrophys. J},
         year = 2002,
        month = mar,
       volume = {568},
       number = {1},
        pages = {289-301},
          doi = {10.1086/338805},
archivePrefix = {arXiv},
       eprint = {astro-ph/0106159},
 primaryClass = {astro-ph},
       adsurl = {https://ui.adsabs.harvard.edu/abs/2002ApJ...568..289A}
}

@article{atnf,
 adsurl               = {http://adsabs.harvard.edu/abs/2005AJ....129.1993M},
 author               = {{Manchester}, R.~N. and {Hobbs}, G.~B. and {Teoh}, A. and {Hobbs}, M.},
 doi                  = {10.1086/428488},
 eprint               = {astro-ph/0412641},
 journal              = {AJ},
 month                = {},
 pages                = {1993-2006},
 title                = {{The Australia Telescope National Facility Pulsar Catalogue}},
 volume               = {129},
 year                 = {2005},
 }

@article{bab17,
 archiveprefix        = {arXiv},
 author               = {{Biryukov}, A. and {Astashenok}, A. and {Beskin}, G.},
 doi                  = {10.1093/mnras/stw3341},
 journal              = {MNRAS},
 pages                = {4320-4331},
 title                = {{Refinement of the timing-bases estimator of pulsars magnetic field}},
 volume               = {466},
 year                 = {2017},
 }

@ARTICLE{bb23,
       author = {{Biryukov}, Anton and {Beskin}, Gregory},
        title = "{Imprint of magnetic obliquity in apparent spin-down of radio pulsars}",
      journal = {MNRAS},
         year = 2023,
        month = jul,
       volume = {522},
       number = {4},
        pages = {6258-6263},
          doi = {10.1093/mnras/stad1437},
archivePrefix = {arXiv},
       eprint = {2305.09184},
 primaryClass = {astro-ph.HE},
       adsurl = {https://ui.adsabs.harvard.edu/abs/2023MNRAS.522.6258B}
}

@ARTICLE{2010MNRAS.403.1829S,
       author = {{Sch{\"o}nrich}, Ralph and {Binney}, James and {Dehnen}, Walter},
        title = "{Local kinematics and the local standard of rest}",
      journal = {MNRAS},
         year = 2010,
        month = apr,
       volume = {403},
       number = {4},
        pages = {1829-1833},
          doi = {10.1111/j.1365-2966.2010.16253.x},
archivePrefix = {arXiv},
       eprint = {0912.3693},
 primaryClass = {astro-ph.GA},
       adsurl = {https://ui.adsabs.harvard.edu/abs/2010MNRAS.403.1829S}
}

@ARTICLE{2024A&A...687A.272D,
       author = {{Disberg}, P. and {Gaspari}, N. and {Levan}, A.~J.},
        title = "{Deceleration of kicked objects due to the Galactic potential}",
      journal = {Astron. and Astrophys.},
         year = 2024,
        month = jul,
       volume = {687},
          eid = {A272},
        pages = {A272},
          doi = {10.1051/0004-6361/202449996},
archivePrefix = {arXiv},
       eprint = {2405.06436},
 primaryClass = {astro-ph.GA},
       adsurl = {https://ui.adsabs.harvard.edu/abs/2024A&A...687A.272D}
}

@ARTICLE{2024Galax..12....7A,
       author = {{Abolmasov}, Pavel and {Biryukov}, Anton and {Popov}, Sergei B.},
        title = "{Spin Evolution of Neutron Stars}",
      journal = {Galaxies},
         year = 2024,
        month = feb,
       volume = {12},
       number = {1},
          eid = {7},
        pages = {7},
          doi = {10.3390/galaxies12010007},
archivePrefix = {arXiv},
       eprint = {2402.04331},
 primaryClass = {astro-ph.HE},
       adsurl = {https://ui.adsabs.harvard.edu/abs/2024Galax..12....7A}
}

@article{bbk07,
 adsurl               = {http://adsabs.harvard.edu/abs/2007AdSpR..40.1498B},
 archiveprefix        = {arXiv},
 author               = {{Biryukov}, A. and {Beskin}, G. and {Karpov}, S. and {Chmyreva}, L.},
 doi                  = {10.1016/j.asr.2007.06.051},
 eprint               = {0709.2549},
 journal              = {Advances in Space Research},
 pages                = {1498-1504},
 title                = {{Evidence of long-term cyclic evolution of radio pulsar periods}},
 volume               = {40},
 year                 = {2007},
 }

@article{bbk12,
 adsurl               = {http://adsabs.harvard.edu/abs/2012MNRAS.420..103B},
 archiveprefix        = {arXiv},
 author               = {{Biryukov}, A. and {Beskin}, G. and {Karpov}, S.},
 doi                  = {10.1111/j.1365-2966.2011.20005.x},
 eprint               = {1105.5019},
 journal              = {MNRAS},
 month                = {feb},
 pages                = {103-117},
 primaryclass         = {astro-ph.HE},
 title                = {{Monotonic and cyclic components of radio pulsar spin-down}},
 volume               = {420},
 year                 = {2012},
 }

@ARTICLE{Verbunt17,
       author = {{Verbunt}, Frank and {Igoshev}, Andrei and {Cator}, Eric},
        title = "{The observed velocity distribution of young pulsars}",
      journal = {Astron. and Astrophys.},
         year = 2017,
        month = dec,
       volume = {608},
          eid = {A57},
        pages = {A57},
          doi = {10.1051/0004-6361/201731518},
archivePrefix = {arXiv},
       eprint = {1708.08281},
 primaryClass = {astro-ph.HE},
       adsurl = {https://ui.adsabs.harvard.edu/abs/2017A&A...608A..57V}
}

@article{car87,
 adsurl               = {http://adsabs.harvard.edu/abs/1987AJ.....94..666C},
 author               = {{Carlberg}, R.~G. and {Innanen}, K.~A.},
 doi                  = {10.1086/114503},
 journal              = {Astron. J.},
 month                = {},
 pages                = {666-670},
 title                = {{Galactic chaos and the circular velocity at the sun}},
 volume               = {94},
 year                 = {1987},
 }

@article{chm10,
 adsurl               = {http://adsabs.harvard.edu/abs/2010AstL...36..116C},
 archiveprefix        = {arXiv},
 author               = {{Chmyreva}, E.~G. and {Beskin}, G.~M. and {Biryukov}, A.~V.},
 doi                  = {10.1134/S1063773710020040},
 eprint               = {1203.2836},
 journal              = {Astronomy Letters},
 month                = {feb},
 pages                = {116-133},
 primaryclass         = {astro-ph.SR},
 title                = {{Search for pairs of isolated radio pulsars{\mdash}Components in disrupted binary systems}},
 volume               = {36},
 year                 = {2010},
 }

@article{edw01,
 adsurl               = {http://adsabs.harvard.edu/abs/2001MNRAS.326..358E},
 author               = {{Edwards}, R.~T. and {Bailes}, M. and {van Straten}, W. and {Britton}, M.~C.},
 doi                  = {10.1046/j.1365-8711.2001.04637.x},
 eprint               = {astro-ph/0105126},
 journal              = {MNRAS},
 month                = {},
 pages                = {358-374},
 title                = {{The Swinburne intermediate-latitude pulsar survey}},
 volume               = {326},
 year                 = {2001},
 }

@article{fgk06,
 adsurl               = {http://adsabs.harvard.edu/abs/2006ApJ...643..332F},
 author               = {{Faucher-Gigu{\`e}re}, C.-A. and {Kaspi}, V.~M.},
 doi                  = {10.1086/501516},
 eprint               = {astro-ph/0512585},
 journal              = {Astrophys. J.},
 month                = {},
 pages                = {332-355},
 title                = {{Birth and Evolution of Isolated Radio Pulsars}},
 volume               = {643},
 year                 = {2006},
 }

@article{gg76,
 adsurl               = {http://adsabs.harvard.edu/abs/1976A%26A....49...57G},
 author               = {{Georgelin}, Y.~M. and {Georgelin}, Y.~P.},
 journal              = {Astron. and Astrophys.},
 month                = {},
 pages                = {57-79},
 title                = {{The spiral structure of our Galaxy determined from H II regions}},
 volume               = {49},
 year                 = {1976},
 }

@article{gullon14,
 adsurl               = {http://adsabs.harvard.edu/abs/2014MNRAS.443.1891G},
 archiveprefix        = {arXiv},
 author               = {{Gull{\'o}n}, M. and {Miralles}, J.~A. and {Vigan{\`o}}, D. and {Pons}, J.~A.},
 doi                  = {10.1093/mnras/stu1253},
 eprint               = {1406.6794},
 journal              = {MNRAS},
 month                = {},
 pages                = {1891-1899},
 primaryclass         = {astro-ph.HE},
 title                = {{Population synthesis of isolated neutron stars with magneto-rotational evolution}},
 volume               = {443},
 year                 = {2014},
 }

@article{has81,
 adsurl               = {http://adsabs.harvard.edu/abs/1981A%26A...100..209H},
 author               = {{Haslam}, C.~G.~T. and {Klein}, U. and {Salter}, C.~J. and {Stoffel}, H. and {Wilson}, W.~E. and {Cleary}, M.~N. and {Cooke}, D.~J. and {Thomasson}, P.},
 journal              = {Astron. and Astrophys.},
 month                = {},
 pages                = {209-219},
 title                = {{A 408 MHz all-sky continuum survey. I - Observations at southern declinations and for the North Polar region}},
 volume               = {100},
 year                 = {1981},
 }

@ARTICLE{2005MNRAS.360..974H,
       author = {{Hobbs}, G. and {Lorimer}, D.~R. and {Lyne}, A.~G. and {Kramer}, M.},
        title = "{A statistical study of 233 pulsar proper motions}",
      journal = {MNRAS},
         year = 2005,
        month = jul,
       volume = {360},
       number = {3},
        pages = {974-992},
          doi = {10.1111/j.1365-2966.2005.09087.x},
archivePrefix = {arXiv},
       eprint = {astro-ph/0504584},
 primaryClass = {astro-ph},
       adsurl = {https://ui.adsabs.harvard.edu/abs/2005MNRAS.360..974H}
}

@article{hobbs04,
 adsurl               = {http://adsabs.harvard.edu/abs/2004MNRAS.353.1311H},
 author               = {{Hobbs}, G. and {Lyne}, A.~G. and {Kramer}, M. and {Martin}, C.~E. and {Jordan}, C.},
 doi                  = {10.1111/j.1365-2966.2004.08157.x},
 journal              = {MNRAS},
 month                = {oct},
 pages                = {1311-1344},
 title                = {{Long-term timing observations of 374 pulsars}},
 volume               = {353},
 year                 = {2004},
 }

@article{hobbs10,
 adsurl               = {http://adsabs.harvard.edu/abs/2010MNRAS.402.1027H},
 archiveprefix        = {arXiv},
 author               = {{Hobbs}, G. and {Lyne}, A.~G. and {Kramer}, M.},
 doi                  = {10.1111/j.1365-2966.2009.15938.x},
 eprint               = {0912.4537},
 journal              = {MNRAS},
 month                = {feb},
 pages                = {1027-1048},
 title                = {{An analysis of the timing irregularities for 366 pulsars}},
 volume               = {402},
 year                 = {2010},
 }

@ARTICLE{ip21,
       author = {{Igoshev}, Andrei P. and {Popov}, Sergei B. and {Hollerbach}, Rainer},
        title = "{Evolution of Neutron Star Magnetic Fields}",
      journal = {Universe},
         year = 2021,
        month = sep,
       volume = {7},
       number = {9},
        pages = {351},
          doi = {10.3390/universe7090351},
archivePrefix = {arXiv},
       eprint = {2109.05584},
 primaryClass = {astro-ph.HE},
       adsurl = {https://ui.adsabs.harvard.edu/abs/2021Univ....7..351I}
}

@article{kui89,
 adsurl               = {http://adsabs.harvard.edu/abs/1989MNRAS.239..571K},
 author               = {{Kuijken}, K. and {Gilmore}, G.},
 doi                  = {10.1093/mnras/239.2.571},
 journal              = {MNRAS},
 month                = {},
 pages                = {571-603},
 title                = {{The mass distribution in the galactic disc. I - A technique to determine the integral surface mass density of the disc near the sun.}},
 volume               = {239},
 year                 = {1989},
 }

@article{law87,
 adsurl               = {http://adsabs.harvard.edu/abs/1987MNRAS.225..307L},
 author               = {{Lawson}, K.~D. and {Mayer}, C.~J. and {Osborne}, J.~L. and {Parkinson}, M.~L.},
 doi                  = {10.1093/mnras/225.2.307},
 journal              = {MNRAS},
 month                = {},
 pages                = {307},
 title                = {{Variations in the Spectral Index of the Galactic Radio Continuum Emission in the Northern Hemisphere}},
 volume               = {225},
 year                 = {1987},
 }

@article{lyne88,
 adsurl               = {http://adsabs.harvard.edu/abs/1988MNRAS.234..477L},
 author               = {{Lyne}, A.~G. and {Manchester}, R.~N.},
 doi                  = {10.1093/mnras/234.3.477},
 journal              = {MNRAS},
 month                = {},
 pages                = {477-508},
 title                = {{The shape of pulsar radio beams}},
 volume               = {234},
 year                 = {1988},
 }

@article{man01,
 adsurl               = {http://adsabs.harvard.edu/abs/2001MNRAS.328...17M},
 author               = {{Manchester}, R.~N. and {Lyne}, A.~G. and {Camilo}, F. and {Bell}, J.~F. and {Kaspi}, V.~M. and {D'Amico}, N. and {McKay}, N.~P.~F. and {Crawford}, F. and {Stairs}, I.~H. and {Possenti}, A. and {Kramer}, M. and {Sheppard}, D.~C.},
 doi                  = {10.1046/j.1365-8711.2001.04751.x},
 eprint               = {astro-ph/0106522},
 journal              = {MNRAS},
 month                = {},
 pages                = {17-35},
 title                = {{The Parkes multi-beam pulsar survey - I. Observing and data analysis systems, discovery and timing of 100 pulsars}},
 volume               = {328},
 year                 = {2001},
 }

@article{nar87,
 adsurl               = {http://adsabs.harvard.edu/abs/1987ApJ...319..162N},
 author               = {{Narayan}, R.},
 doi                  = {10.1086/165442},
 journal              = {Astrophys. J.},
 month                = {},
 pages                = {162-179},
 title                = {{The birthrate and initial spin period of single radio pulsars}},
 volume               = {319},
 year                 = {1987},
 }

@ARTICLE{maciesiak11a,
       author = {{Maciesiak}, Krzysztof and {Gil}, Janusz and {Ribeiro}, Val{\'e}rio A.~R.~M.},
        title = "{On the pulse-width statistics in radio pulsars - I. Importance of the interpulse emission}",
      journal = {MNRAS},
         year = 2011,
        month = jun,
       volume = {414},
       number = {2},
        pages = {1314-1328},
          doi = {10.1111/j.1365-2966.2011.18471.x},
archivePrefix = {arXiv},
       eprint = {1102.3348},
 primaryClass = {astro-ph.GA},
       adsurl = {https://ui.adsabs.harvard.edu/abs/2011MNRAS.414.1314M}
}

@ARTICLE{malov13,
   author = {{Malov}, I.~F. and {Nikitina}, E.~B.},
    title = "{The magnetospheric structure of radio pulsars with interpulses}",
  journal = {Astronomy Reports},
     year = 2013,
    month = nov,
   volume = 57,
    pages = {833-843},
      doi = {10.1134/S106377291311005X},
   adsurl = {https://ui.adsabs.harvard.edu/abs/2013ARep...57..833M}
}

@ARTICLE{Johnston05,
       author = {{Johnston}, Simon and {Hobbs}, G. and {Vigeland}, S. and {Kramer}, M. and {Weisberg}, J.~M. and {Lyne}, A.~G.},
        title = "{Evidence for alignment of the rotation and velocity vectors in pulsars}",
      journal = {MNRAS},
         year = 2005,
        month = dec,
       volume = {364},
       number = {4},
        pages = {1397-1412},
          doi = {10.1111/j.1365-2966.2005.09669.x},
archivePrefix = {arXiv},
       eprint = {astro-ph/0510260},
 primaryClass = {astro-ph},
       adsurl = {https://ui.adsabs.harvard.edu/abs/2005MNRAS.364.1397J}
}

@ARTICLE{Brisken02,
       author = {{Brisken}, Walter F. and {Benson}, John M. and {Goss}, W.~M. and {Thorsett}, S.~E.},
        title = "{Very Long Baseline Array Measurement of Nine Pulsar Parallaxes}",
      journal = {Astrophys. J},
         year = 2002,
        month = jun,
       volume = {571},
       number = {2},
        pages = {906-917},
          doi = {10.1086/340098},
archivePrefix = {arXiv},
       eprint = {astro-ph/0204105},
 primaryClass = {astro-ph},
       adsurl = {https://ui.adsabs.harvard.edu/abs/2002ApJ...571..906B}
}

@ARTICLE{Brisken03,
       author = {{Brisken}, W.~F. and {Fruchter}, A.~S. and {Goss}, W.~M. and {Herrnstein}, R.~M. and {Thorsett}, S.~E.},
        title = "{Proper-Motion Measurements with the VLA. II. Observations of 28 Pulsars}",
      journal = {Astron. J},
         year = 2003,
        month = dec,
       volume = {126},
       number = {6},
        pages = {3090-3098},
          doi = {10.1086/379559},
archivePrefix = {arXiv},
       eprint = {astro-ph/0309215},
 primaryClass = {astro-ph},
       adsurl = {https://ui.adsabs.harvard.edu/abs/2003AJ....126.3090B}
}

@ARTICLE{Meerkat_RVM,
       author = {{Johnston}, S. and {Kramer}, M. and {Karastergiou}, A. and {Keith}, M.~J. and {Oswald}, L.~S. and {Parthasarathy}, A. and {Weltevrede}, P.},
        title = "{The Thousand-Pulsar-Array programme on MeerKAT - XI. Application of the rotating vector model}",
      journal = {MNRAS},
         year = 2023,
        month = apr,
       volume = {520},
       number = {4},
        pages = {4801-4814},
          doi = {10.1093/mnras/stac3636},
archivePrefix = {arXiv},
       eprint = {2212.03988},
 primaryClass = {astro-ph.HE},
       adsurl = {https://ui.adsabs.harvard.edu/abs/2023MNRAS.520.4801J}
}

@article{nout13,
 adsurl               = {http://adsabs.harvard.edu/abs/2013MNRAS.430.2281N},
 archiveprefix        = {arXiv},
 author               = {{Noutsos}, A. and {Schnitzeler}, D.~H.~F.~M. and {Keane}, E.~F. and {Kramer}, M. and {Johnston}, S.},
 doi                  = {10.1093/mnras/stt047},
 eprint               = {1301.1265},
 journal              = {MNRAS},
 month                = {},
 pages                = {2281-2301},
 title                = {{Pulsar spin-velocity alignment: kinematic ages, birth periods and braking indices}},
 volume               = {430},
 year                 = {2013},
 }

@ARTICLE{nout12,
       author = {{Noutsos}, A. and {Kramer}, M. and {Carr}, P. and {Johnston}, S.},
        title = "{Pulsar spin-velocity alignment: further results and discussion}",
      journal = {MNRAS},
         year = 2012,
        month = jul,
       volume = {423},
       number = {3},
        pages = {2736-2752},
          doi = {10.1111/j.1365-2966.2012.21083.x},
archivePrefix = {arXiv},
       eprint = {1205.2305},
 primaryClass = {astro-ph.GA},
       adsurl = {https://ui.adsabs.harvard.edu/abs/2012MNRAS.423.2736N}
}

@article{phil14,
 adsurl               = {http://adsabs.harvard.edu/abs/2014MNRAS.441.1879P},
 archiveprefix        = {arXiv},
 author               = {{Philippov}, A. and {Tchekhovskoy}, A. and {Li}, J.~G.},
 doi                  = {10.1093/mnras/stu591},
 eprint               = {1311.1513},
 journal              = {MNRAS},
 month                = {},
 pages                = {1879-1887},
 primaryclass         = {astro-ph.HE},
 title                = {{Time evolution of pulsar obliquity angle from 3D simulations of magnetospheres}},
 volume               = {441},
 year                 = {2014},
 }

@article{rankin93a,
 adsurl               = {http://adsabs.harvard.edu/abs/1993ApJS...85..145R},
 author               = {{Rankin}, J.~M.},
 doi                  = {10.1086/191758},
 journal              = {Astrophys. J. Suppl.},
 month                = {},
 pages                = {145-161},
 title                = {{Toward an empirical theory of pulsar emission. VI - The geometry of the conal emission region: Appendix and tables}},
 volume               = {85},
 year                 = {1993},
 }

@ARTICLE{rankin90,
       author = {{Rankin}, Joanna M.},
        title = "{Toward an Empirical Theory of Pulsar Emission. IV. Geometry of the Core Emission Region}",
      journal = {Astrophys. J},
         year = 1990,
        month = mar,
       volume = {352},
        pages = {247},
          doi = {10.1086/168530},
       adsurl = {https://ui.adsabs.harvard.edu/abs/1990ApJ...352..247R}
}

@article{rankin93b,
 adsurl               = {http://adsabs.harvard.edu/abs/1993ApJ...405..285R},
 author               = {{Rankin}, J.~M.},
 doi                  = {10.1086/172361},
 journal              = {Astrophys. J.},
 month                = {},
 pages                = {285-297},
 title                = {{Toward an empirical theory of pulsar emission. VI - The geometry of the conal emission region}},
 volume               = {405},
 year                 = {1993},
 }

@article{raw86,
 adsurl               = {http://adsabs.harvard.edu/abs/1986Natur.319..383R},
 author               = {{Rawley}, L.~A. and {Taylor}, J.~H. and {Davis}, M.~M.},
 doi                  = {10.1038/319383a0},
 journal              = {Nature},
 month                = {},
 pages                = {383},
 title                = {{Period derivative and orbital eccentricity of binary pulsar 1953 + 29}},
 volume               = {319},
 year                 = {1986},
 }

@ARTICLE{Tetzlaff_HyperVel11,
       author = {{Tetzlaff}, N. and {Neuh{\"a}user}, R. and {Hohle}, M.~M.},
        title = "{A catalogue of young runaway Hipparcos stars within 3 kpc from the Sun}",
      journal = {MNRAS},
         year = 2011,
        month = jan,
       volume = {410},
       number = {1},
        pages = {190-200},
          doi = {10.1111/j.1365-2966.2010.17434.x},
archivePrefix = {arXiv},
       eprint = {1007.4883},
 primaryClass = {astro-ph.GA},
       adsurl = {https://ui.adsabs.harvard.edu/abs/2011MNRAS.410..190T}
}

@ARTICLE{Carre_HyperVel23,
       author = {{Carretero-Castrillo}, M. and {Rib{\'o}}, M. and {Paredes}, J.~M.},
        title = "{Galactic runaway O and Be stars found using Gaia DR3}",
      journal = {Astron. and Astrophys.},
         year = 2023,
        month = nov,
       volume = {679},
          eid = {A109},
        pages = {A109},
          doi = {10.1051/0004-6361/202346613},
archivePrefix = {arXiv},
       eprint = {2311.01827},
 primaryClass = {astro-ph.SR},
       adsurl = {https://ui.adsabs.harvard.edu/abs/2023A&A...679A.109C}
}

@ARTICLE{1997A&A...327..155G,
       author = {{Gangadhara}, R.~T.},
        title = "{Orthogonal polarization mode phenomenon in pulsars.}",
      journal = {Astron. and Astrophys.},
         year = 1997,
        month = nov,
       volume = {327},
        pages = {155-166},
          doi = {10.48550/arXiv.astro-ph/9707168},
archivePrefix = {arXiv},
       eprint = {astro-ph/9707168},
 primaryClass = {astro-ph},
       adsurl = {https://ui.adsabs.harvard.edu/abs/1997A&A...327..155G}
}

@ARTICLE{2014MNRAS.441.1943W,
       author = {{Wang}, P.~F. and {Wang}, C. and {Han}, J.~L.},
        title = "{Polarized curvature radiation in pulsar magnetosphere}",
      journal = {MNRAS},
         year = 2014,
        month = jul,
       volume = {441},
       number = {3},
        pages = {1943-1953},
          doi = {10.1093/mnras/stu690},
archivePrefix = {arXiv},
       eprint = {1404.1431},
 primaryClass = {astro-ph.HE},
       adsurl = {https://ui.adsabs.harvard.edu/abs/2014MNRAS.441.1943W}
}

@ARTICLE{1984ApJS...55..247S,
       author = {{Stinebring}, D.~R. and {Cordes}, J.~M. and {Rankin}, J.~M. and {Weisberg}, J.~M. and {Boriakoff}, V.},
        title = "{Pulsar polarization fluctuations. I. 1404 MHz statistical summaries.}",
      journal = {Astrophys. J. Suppl.},
         year = 1984,
        month = jun,
       volume = {55},
        pages = {247-277},
          doi = {10.1086/190954},
       adsurl = {https://ui.adsabs.harvard.edu/abs/1984ApJS...55..247S}
}

@ARTICLE{Mandel_Igoshev23,
       author = {{Mandel}, Ilya and {Igoshev}, Andrei P.},
        title = "{The Impact of Spin-kick Alignment on the Inferred Velocity Distribution of Isolated Pulsars}",
      journal = {Astrophys. J},
         year = 2023,
        month = feb,
       volume = {944},
       number = {2},
          eid = {153},
        pages = {153},
          doi = {10.3847/1538-4357/acb3c3},
archivePrefix = {arXiv},
       eprint = {2210.12305},
 primaryClass = {astro-ph.HE},
       adsurl = {https://ui.adsabs.harvard.edu/abs/2023ApJ...944..153M}
}

@article{rs75,
 adsurl               = {http://adsabs.harvard.edu/abs/1975ApJ...196...51R},
 author               = {{Ruderman}, M.~A. and {Sutherland}, P.~G.},
 doi                  = {10.1086/153393},
 journal              = {Astrophys. J.},
 month                = {},
 pages                = {51-72},
 title                = {{Theory of pulsars - Polar caps, sparks, and coherent microwave radiation}},
 volume               = {196},
 year                 = {1975},
 }

@ARTICLE{jk19,
       author = {{Johnston}, Simon and {Kramer}, Michael},
        title = "{On the beam properties of radio pulsars with interpulse emission}",
      journal = {MNRAS},
         year = 2019,
        month = dec,
       volume = {490},
       number = {4},
        pages = {4565-4574},
          doi = {10.1093/mnras/stz2865},
archivePrefix = {arXiv},
       eprint = {1910.04550},
 primaryClass = {astro-ph.HE},
       adsurl = {https://ui.adsabs.harvard.edu/abs/2019MNRAS.490.4565J}
}

@ARTICLE{keith10,
       author = {{Keith}, M.~J. and {Johnston}, S. and {Weltevrede}, P. and {Kramer}, M.},
        title = "{Polarization measurements of five pulsars with interpulses}",
      journal = {MNRAS},
         year = 2010,
        month = feb,
       volume = {402},
       number = {2},
        pages = {745-752},
          doi = {10.1111/j.1365-2966.2009.15926.x},
archivePrefix = {arXiv},
       eprint = {0910.4778},
 primaryClass = {astro-ph.SR},
       adsurl = {https://ui.adsabs.harvard.edu/abs/2010MNRAS.402..745K}
}

@ARTICLE{RVM,
       author = {{Radhakrishnan}, V. and {Cooke}, D.~J.},
        title = "{Magnetic Poles and the Polarization Structure of Pulsar Radiation}",
      journal = {Astrophys. J. Lett.},
         year = 1969,
        month = jan,
       volume = {3},
        pages = {225},
       adsurl = {https://ui.adsabs.harvard.edu/abs/1969ApL.....3..225R}
}

@ARTICLE{TEMPO2,
       author = {{Edwards}, R.~T. and {Hobbs}, G.~B. and {Manchester}, R.~N.},
        title = "{TEMPO2, a new pulsar timing package - II. The timing model and precision estimates}",
      journal = {MNRAS},
         year = 2006,
        month = nov,
       volume = {372},
       number = {4},
        pages = {1549-1574},
          doi = {10.1111/j.1365-2966.2006.10870.x},
archivePrefix = {arXiv},
       eprint = {astro-ph/0607664},
 primaryClass = {astro-ph},
       adsurl = {https://ui.adsabs.harvard.edu/abs/2006MNRAS.372.1549E}
}

@ARTICLE{Wang06_SpinVel,
       author = {{Wang}, Chen and {Lai}, Dong and {Han}, J.~L.},
        title = "{Neutron Star Kicks in Isolated and Binary Pulsars: Observational Constraints and Implications for Kick Mechanisms}",
      journal = {Astrophys. J},
         year = 2006,
        month = mar,
       volume = {639},
       number = {2},
        pages = {1007-1017},
          doi = {10.1086/499397},
archivePrefix = {arXiv},
       eprint = {astro-ph/0509484},
 primaryClass = {astro-ph},
       adsurl = {https://ui.adsabs.harvard.edu/abs/2006ApJ...639.1007W}
}

@ARTICLE{Wang07_SpinVel,
       author = {{Wang}, Chen and {Lai}, Dong and {Han}, J.~L.},
        title = "{Spin-Kick Correlation in Neutron Stars: Alignment Conditions and Implications}",
      journal = {Astrophys. J},
         year = 2007,
        month = feb,
       volume = {656},
       number = {1},
        pages = {399-407},
          doi = {10.1086/510352},
archivePrefix = {arXiv},
       eprint = {astro-ph/0607666},
 primaryClass = {astro-ph},
       adsurl = {https://ui.adsabs.harvard.edu/abs/2007ApJ...656..399W}
}

@article{spitkovsky06,
 adsurl               = {http://adsabs.harvard.edu/abs/2006ApJ...648L..51S},
 author               = {{Spitkovsky}, A.},
 doi                  = {10.1086/507518},
 eprint               = {astro-ph/0603147},
 journal              = {Astrophys. J. Lett.},
 month                = {},
 pages                = {L51-L54},
 title                = {{Time-dependent Force-free Pulsar Magnetospheres: Axisymmetric and Oblique Rotators}},
 volume               = {648},
 year                 = {2006},
 }

@ARTICLE{Yao21,
       author = {{Yao}, Jumei and {Zhu}, Weiwei and {Manchester}, Richard N. and {Coles}, William A. and {Li}, Di and {Wang}, Na and {Kramer}, Michael and {Stinebring}, Daniel R. and {Feng}, Yi and {Yan}, Wenming and {Miao}, Chenchen and {Yuan}, Mao and {Wang}, Pei and {Lu}, Jiguang},
        title = "{Evidence for three-dimensional spin-velocity alignment in a pulsar}",
      journal = {Nature Astronomy},
         year = 2021,
        month = may,
       volume = {5},
        pages = {788-795},
          doi = {10.1038/s41550-021-01360-w},
archivePrefix = {arXiv},
       eprint = {2103.01839},
 primaryClass = {astro-ph.GA},
       adsurl = {https://ui.adsabs.harvard.edu/abs/2021NatAs...5..788Y}
}

@article{wai92,
 adsurl               = {http://adsabs.harvard.edu/abs/1992ApJS...83..111W},
 author               = {{Wainscoat}, R.~J. and {Cohen}, M. and {Volk}, K. and {Walker}, H.~J. and {Schwartz}, D.~E.},
 doi                  = {10.1086/191733},
 journal              = {Astrophys. J. Suppl.},
 month                = {},
 pages                = {111-146},
 title                = {{A model of the 8-25 micron point source infrared sky}},
 volume               = {83},
 year                 = {1992},
 }

@article{WFF,
 adsurl               = {http://adsabs.harvard.edu/abs/1988PhRvC..38.1010W},
 author               = {{Wiringa}, R.~B. and {Fiks}, V. and {Fabrocini}, A.},
 doi                  = {10.1103/PhysRevC.38.1010},
 journal              = {Phys. Rev. C},
 month                = {},
 pages                = {1010-1037},
 title                = {{Equation of state for dense nucleon matter}},
 volume               = {38},
 year                 = {1988},
 }

@article{yk04,
 adsurl               = {http://adsabs.harvard.edu/abs/2004A%26A...422..545Y},
 author               = {{Yusifov}, I. and {K{\"u}{\c c}{\"u}k}, I.},
 doi                  = {10.1051/0004-6361:20040152},
 eprint               = {astro-ph/0405559},
 journal              = {Astron. and Astrophys.},
 month                = {},
 pages                = {545-553},
 title                = {{Revisiting the radial distribution of pulsars in the Galaxy}},
 volume               = {422},
 year                 = {2004},
 }

@ARTICLE{Janka17,
       author = {{Janka}, Hans-Thomas},
        title = "{Neutron Star Kicks by the Gravitational Tug-boat Mechanism in Asymmetric Supernova Explosions: Progenitor and Explosion Dependence}",
      journal = {Astrophys. J},
         year = 2017,
        month = mar,
       volume = {837},
       number = {1},
          eid = {84},
        pages = {84},
          doi = {10.3847/1538-4357/aa618e},
archivePrefix = {arXiv},
       eprint = {1611.07562},
 primaryClass = {astro-ph.HE},
       adsurl = {https://ui.adsabs.harvard.edu/abs/2017ApJ...837...84J}
}

@ARTICLE{Janka22,
       author = {{Janka}, Hans-Thomas and {Wongwathanarat}, Annop and {Kramer}, Michael},
        title = "{Supernova Fallback as Origin of Neutron Star Spins and Spin-kick Alignment}",
      journal = {Astrophys. J},
         year = 2022,
        month = feb,
       volume = {926},
       number = {1},
          eid = {9},
        pages = {9},
          doi = {10.3847/1538-4357/ac403c},
archivePrefix = {arXiv},
       eprint = {2104.07493},
 primaryClass = {astro-ph.HE},
       adsurl = {https://ui.adsabs.harvard.edu/abs/2022ApJ...926....9J}
}

@ARTICLE{RVM_Lyutikov,
       author = {{Lyutikov}, Maxim},
        title = "{Relativistic Rotating Vector Model}",
      journal = {arXiv e-prints},
         year = 2016,
        month = jul,
          eid = {arXiv:1607.00777},
        pages = {arXiv:1607.00777},
          doi = {10.48550/arXiv.1607.00777},
archivePrefix = {arXiv},
       eprint = {1607.00777},
 primaryClass = {astro-ph.HE},
       adsurl = {https://ui.adsabs.harvard.edu/abs/2016arXiv160700777L}
}

@ARTICLE{FAST_polarization23,
       author = {{Wang}, P.~F. and {Han}, J.~L. and {Xu}, J. and {Wang}, C. and {Yan}, Y. and {Jing}, W.~C. and {Su}, W.~Q. and {Zhou}, D.~J. and {Wang}, T.},
        title = "{FAST Pulsar Database. I. Polarization Profiles of 682 Pulsars}",
      journal = {Research in Astronomy and Astrophysics},
         year = 2023,
        month = oct,
       volume = {23},
       number = {10},
          eid = {104002},
        pages = {104002},
          doi = {10.1088/1674-4527/acea1f},
archivePrefix = {arXiv},
       eprint = {2307.10340},
 primaryClass = {astro-ph.HE},
       adsurl = {https://ui.adsabs.harvard.edu/abs/2023RAA....23j4002W}
}

@ARTICLE{Igoshev20,
       author = {{Igoshev}, Andrei P.},
        title = "{The observed velocity distribution of young pulsars - II. Analysis of complete PSR{\ensuremath{\pi}}}",
      journal = {MNRAS},
         year = 2020,
        month = may,
       volume = {494},
       number = {3},
        pages = {3663-3674},
          doi = {10.1093/mnras/staa958},
archivePrefix = {arXiv},
       eprint = {2002.01367},
 primaryClass = {astro-ph.HE},
       adsurl = {https://ui.adsabs.harvard.edu/abs/2020MNRAS.494.3663I}
}

@ARTICLE{Igoshev22,
       author = {{Igoshev}, Andrei P. and {Frantsuzova}, Anastasia and {Gourgouliatos}, Konstantinos N. and {Tsichli}, Savina and {Konstantinou}, Lydia and {Popov}, Sergei B.},
        title = "{Initial periods and magnetic fields of neutron stars}",
      journal = {MNRAS},
         year = 2022,
        month = aug,
       volume = {514},
       number = {3},
        pages = {4606-4619},
          doi = {10.1093/mnras/stac1648},
archivePrefix = {arXiv},
       eprint = {2205.06823},
 primaryClass = {astro-ph.HE},
       adsurl = {https://ui.adsabs.harvard.edu/abs/2022MNRAS.514.4606I}
}

\appendix

\section{Analyzed pulsar subset details and expanded results of the population synthesis}
\label{app:pop_synth_results}

\begin{table*}
\begin{threeparttable}
\caption{Pulsars under consideration. These classical isolated pulsars are assumed to be either weakly or strongly inclined in terms of their magnetic angle. At the same time, they all have distance estimates and full proper motions. The ages here are characteristic ages $P/2\dot P$ in $10^6$ years, while the surface fields are magnetodipolar estimates of $3.2\times 10^{19} \sqrt{P \dot P}$ in $10^{12}$ Gs. The last column lists the references where a particular pulsar has been found to be either aligned or orthogonal.}
\label{tab:pulsars}
\begin{tabular}{rlrrrrrrll}
\toprule
\headrow N & PSR & $P$, sec & Age, Myr & $B_{12}$, Gs & Distance, kpc & $v_{\mathrm t}$, km s$^{-1}$ & $v_{\mathrm pec, t}$, km s$^{-1}$ & Type & References \\
\midrule
1  &   J0152-1637   &   0.833   &  10.17  &    1.05  &    2.00 (PX)\tnote{a}  &   206 &   209 & orthogonal   & [1], [3]\tnote{b} \\
\midrule
2  &   J0406+6138   &   0.595   &   1.69  &    1.84  &    4.55 (PX)  &   558 &   561 & orthogonal   &  [3] \\ 
\midrule
3  &   J0525+1115   &   0.354   &  76.44  &    0.16  &    1.84 (DM)  &   267 &   262 & orthogonal   & [2] \\
\midrule
4  &   J0534+2200   &   0.033   &   0.00  &    3.79  &    2.00 (DM)  &   143 &   152 & orthogonal   &  [2], [4] \\
\midrule
5  &   J0820-1350   &   1.238   &   9.31  &    1.64  &    1.96 (PX)  &   437 &   464 & orthogonal   &  [1] \\
\midrule
6  &   J0826+2637   &   0.531   &   4.93  &    0.96  &    0.50 (PX)  &   235 &   232 & orthogonal   &  [1], [2], [4] \\
\midrule
7  &   J0835-4510   &   0.089   &   0.01  &    3.38  &    0.29 (PX)  &     62&     45& orthogonal   &  [1], [2] \\
\midrule
8  &   J0908-4913   &   0.107   &   0.11  &    1.28  &    1.00 (DM)  &   193 &   156 & orthogonal   &  [2], [4], [5], [6], [7] \\
\midrule
9  &   J1057-5226   &   0.197   &   0.54  &    1.09  &    0.09 (DM)  &    34 &    52 & orthogonal   &  [1], [2], [4], [5]  \\
\midrule
10 &   J1509+5531   &   0.740   &   2.35  &    1.95  &    2.13 (PX)  &   964 &   955 & orthogonal   &  [1] \\
\midrule
11 &   J1645-0317   &   0.388   &   3.46  &    0.84  &    3.85 (PX)  &   374 &   402 & orthogonal   &  [1], []2 \\
\midrule
12 &   J1705-1906   &   0.299   &   1.15  &    1.13  &    0.75 (DM)  &   314 &   306 & orthogonal   &  [1], [2], [4], [6] \\
\midrule
13 &   J1722-3207   &   0.477   &  11.74  &    0.56  &    2.93 (DM)  &   541 &   524 & orthogonal   & [1] \\
\midrule
14 &   J1731-4744   &   0.830   &   0.08  &   11.81  &    0.70 (DM)  &  485  &  470  & orthogonal   &  [1] \\
\midrule
15 &   J1751-4657   &   0.742   &   9.06  &    0.99  &    0.74 (DM)  &  193  &  176  & orthogonal   &  [1] \\
\midrule
16 &   J1820-0427   &   0.598   &   1.50  &    1.97  &    2.86 (PX)  &  262  &  290  & orthogonal   &  [2] \\
\midrule
17 &   J1841+0912   &   0.381   &   5.55  &    0.65  &    1.66 (DM)  &   337 &   361 & orthogonal   &  [1], [3] \\
\midrule
18 &   J1903+0135   &   0.729   &   2.87  &    1.73  &    3.30 (DM)  &   167 &   133 & orthogonal   & [1] \\
\midrule
19 &   J1909+0007   &   1.017   &   2.92  &    2.40  &    4.36 (DM)  &   602 &   539 & orthogonal   &  [2], [3] \\
\midrule
20 &   J1909+1102   &   0.284   &   1.71  &    0.88  &    4.80 (DM)  &   243 &   314 & orthogonal   & [1] \\
\midrule
21 &   J1913-0440   &   0.826   &   3.22  &    1.86  &    4.04 (DM)  &   167 &   191 & orthogonal   &  [1] \\
\midrule
22 &   J1917+1353   &   0.195   &   0.43  &    1.20  &    5.88 (PX)  &   260 &   413 & orthogonal   & [2] \\
\midrule
23 &   J1919+0021   &   1.272   &   2.63  &    3.16  &    5.88 (PX)  &   345 &   425 & orthogonal   &  [1], [2] \\
\midrule
24 &   J2022+2854   &   0.343   &   2.88  &    0.82  &    2.70 (PX)  &  246  &  187  & orthogonal   & [1], [2] \\
\midrule
25 &   J2330-2005   &   1.644   &   5.63  &    2.79  &    0.86 (DM)  &  262  &  249  & orthogonal   & [2] \\
\midrule
\midrule
1   &   J0157+6212  &   2.352  &    0.20  &   21.33   &   1.79 (PX)  &     384   &    391  &  aligned    &  [2] \\
\midrule
2   &   J0502+4654  &   0.639  &    1.82  &    1.91   &   1.32 (DM)  &      80   &     96	&  aligned   &  [1], [3] \\
\midrule
3   &   J0659+1414  &   0.385  &    0.11  &    4.65   &   0.29 (PX)  &      65   &     73	&	aligned  & [1] \\
\midrule
4   &   J0946+0951  &   1.098  &    4.99  &    1.98   &   0.89 (DM)  &    163    &   141	&	aligned  & [1] \\
\midrule
5   &   J0953+0755  &   0.253  &   17.46  &    0.24   &   0.26 (PX)  &     39    &    55	&	aligned  & [1], [4], [6] \\
\midrule
6   &   J1302-6350  &   0.048  &    0.33  &    0.33   &   2.63 (PX)  &     14    &    63	&	aligned  & [4]  \\
\midrule
7   &   J1543+0929  &   0.748  &   27.50  &    0.58   &   7.69 (PX)  &    269    &   293	&	aligned  &   [1], [2] \\
\midrule
8   &   J1720-0212  &   0.478  &   91.58  &    0.20   &   2.36 (DM)  &     259   &    238	&	aligned  &  [1] \\
\midrule
9  &   J1946+1805  &   0.441  &  290.20  &    0.10   &   0.30 (DM)  &       6   &      8	&	aligned  & [4], [6]  \\
\midrule
10  &   J2006-0807  &   0.581  &  200.44  &    0.17   &   2.63 (PX)  &    114    &    80	&	aligned  &  [1], [3]\\
\midrule
11  &   J2113+4644  &   1.015  &   22.53  &    0.86   &   2.17 (PX)  &    181    &   226	&	aligned  &  [1], [2] \\
\midrule
12  &   J2149+6329  &   0.380  &   35.92  &    0.26   &   2.78 (PX)  &    299    &   343	&	aligned  & [1]  \\
\midrule
13  &   J2325+6316  &   1.436  &    8.06  &    2.04   &   4.86 (DM)  &     83    &    33	&	aligned  &  [1] \\
\bottomrule
\end{tabular}
\begin{tablenotes}[hang]
\item[a]DM is for dispersion measure-based distances according to Galactic free electrons distribution by [REF]; PX for the parallax-based distances.
\item[b]References that support the type of the obliquity: [1] \cite{lyne88}; [2] \cite{rankin90}; [3] \cite{rankin93a}; 
 [4] \cite{maciesiak11a}; [5] \cite{keith10}; [6] \cite{malov13}; [7] \cite{jk19}.
%\item[a]First note
%\item[b]Another table note
\end{tablenotes}
\end{threeparttable}
\end{table*}

\begin{figure*}
\centering
\includegraphics[width=1\linewidth]{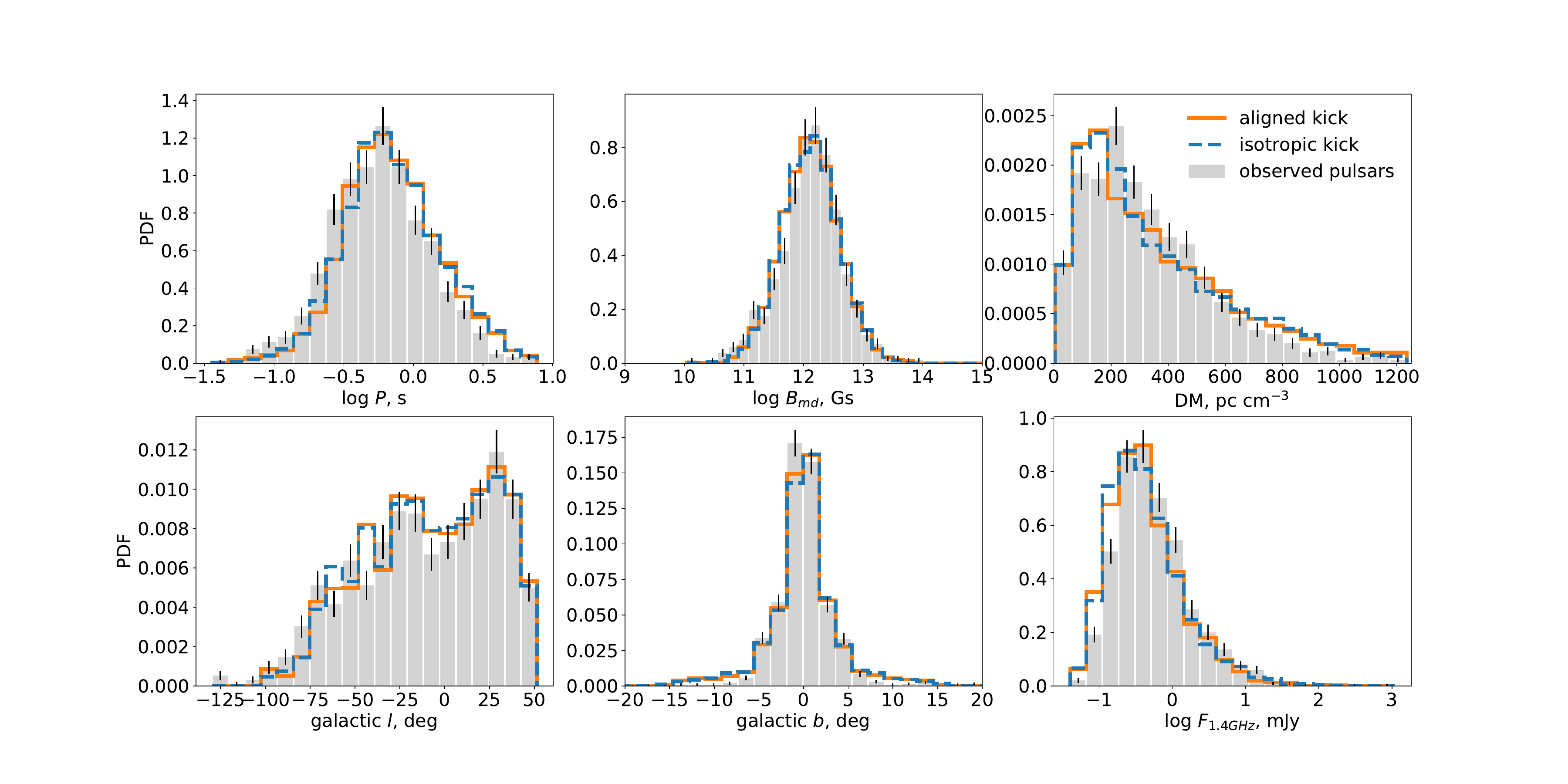}
\caption{Pulsar population synthesis results show distributions of six observables used to find optimal model parameters: pulsar spin period $P$, magnetic field $B_{\mathrm md} \propto \sqrt{P \dot{P}}$, dispersion measure $DM$, galactic coordinates $(l,b)$, and flux $F$ at 1.4 GHz. The solid orange line shows synthetic distributions assuming a kick velocity aligned with the spin axis, while the dashed blue line represents an isotropic kick. Both models use the same initial parameters from Table~\ref{tab:synthesis_results}. Observed distributions for 1057 single classical pulsars from Parkes and Swinburne surveys are shown as grey bars. Population synthesis qualitatively reproduces the observed statistics well, and the plotted quantities are insensitive to the possible spin-kick correlation.}
\label{fig:synthesis_results}
\end{figure*}

\begin{figure*}
\centering
\includegraphics[width=1\linewidth]{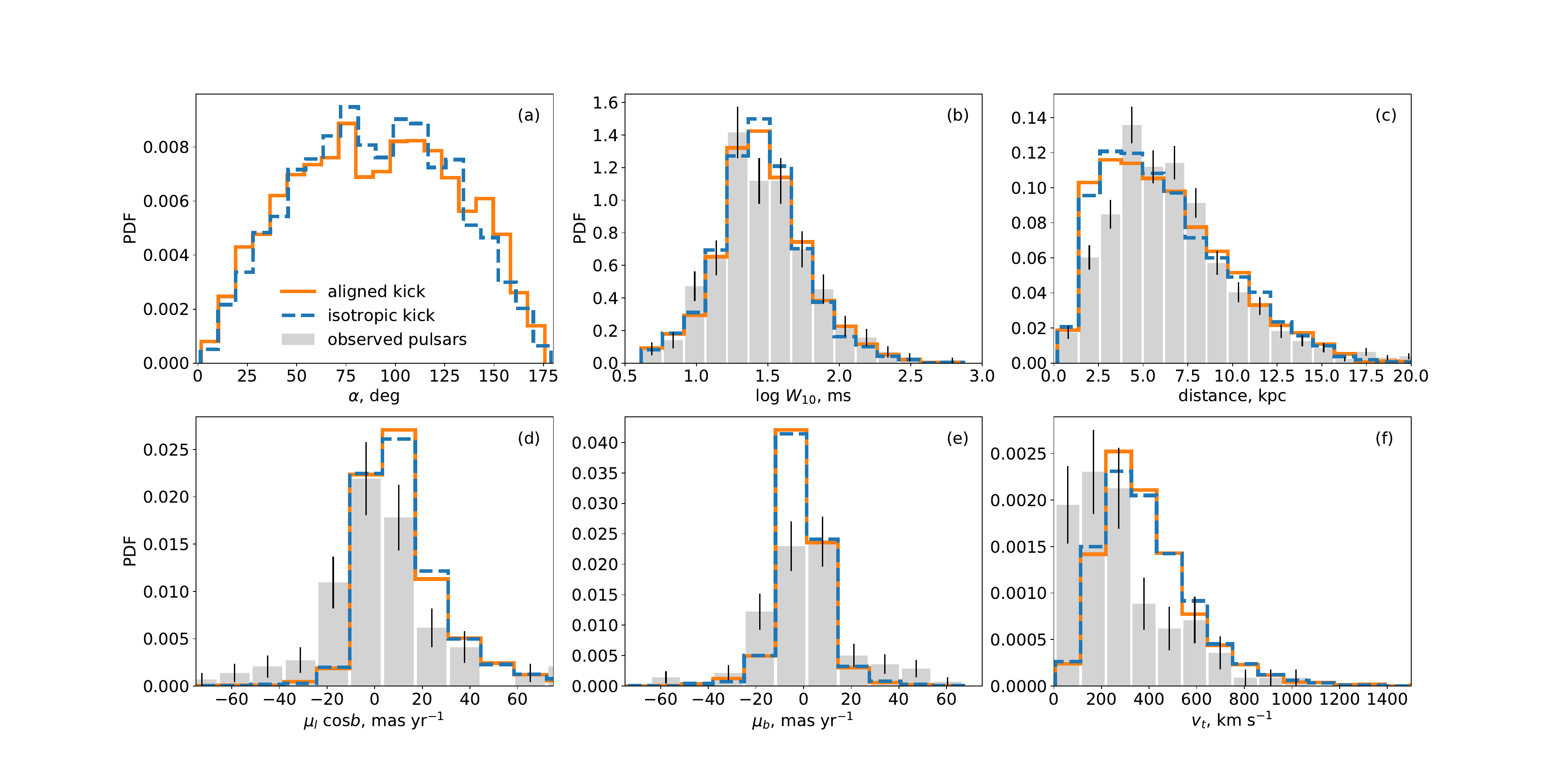}
\caption{Further results of population synthesis. This figure is organised similarly to the previous one: the grey-shaded bars represent the observed distributions (where applicable), while the orange solid and blue dashed lines are for synthetic distributions. {\bf (a)} Magnetic angle $\alpha$ between the spin and magnetic angle of synthetic pulsars. Surprisingly, this parameter is insensitive to the specific spin-kick orientation scenario. This is probably due to the nearly anisotropic distribution of pulsars relative to the observer. {\bf (b)}  Pulsar pulse widths at 10\% of maximum in miliseconds. The observed distribution is shown for 381 pulsars with measured $W_{10}$ from the control subset of 1057 objects. {\bf (c)} Distances to pulsars relative to the Solar System barycenter. Observed values are mostly based on the dispersion measure. Here, all 1057 control pulsars are shown. {\bf (d)} Proper motion along the galactic longitude. The observed distribution is shown for 106 pulsars with known $\mu_l$ from the initial control subset. {\bf (e)} Similar, but for galactic {\bf latitudinal} direction. {\bf (f)} Transverse velocities relative to the Solar system barycenter. Relative to this particular control subset of pulsars, the synthetic population shows an excess in nearby and faster pulsars. This could be a result of a combination of selection effects in the estimation of pulsar proper motions, systematic errors in estimations of distances, as well as some incompleteness of the kick distribution model.}
\label{fig:synth_derivs}
\end{figure*}

\end{document}